\documentclass[seceqn,11pt]{elsart}

\usepackage{amsmath,amssymb,latexsym,amssymb,tables,psfig,mathrsfs}
\usepackage{macros,subfigure,psfrag,color,bbm}
\usepackage[font=scriptsize,labelfont=bf]{caption}

\usepackage[dvips]{epsfig}
\usepackage[dvips]{graphicx}
\usepackage{longtable,rotating}

\input eqalign.sty

\begin{document}

\begin{frontmatter}

\begin{flushright}
DESY 12-137  \\
HU-EP-12/24 \\
JLAB-THY-12-1606 \\
SFB/CPP-12-55 \\
\end{flushright}

\title{A quenched study of the Schr\"odinger functional 
with chirally rotated boundary conditions: applications} 

\author[HU,DESY]{J. Gonz\'alez L\'opez},
\author[DESY]{K. Jansen},
\author[JLAB]{D.\ B.\ Renner},
\author[HU]{A. Shindler}\footnote{Heisenberg Fellow}
\address[HU]{Humboldt-Universit\"at zu Berlin, Institut f\"ur Physik
  Newtonstrasse 15, 12489 Berlin, Germany}
\address[DESY]{DESY,Platanenallee 6, 15738 Zeuthen, Germany}
\address[JLAB]{Jefferson Lab, 12000 Jefferson Avenue, Newport News, VA 23606, USA}

\maketitle
\begin{abstract}
In a previous paper~\cite{Lopez:2012as}, we have discussed the non-perturbative tuning of the 
chirally rotated Schr\"odinger functional ($\chi$SF).
This tuning is required to eliminate bulk O($a$) cutoff effects in physical correlation functions.
Using our tuning results obtained in~\cite{Lopez:2012as} we 
perform scaling and universality tests analyzing the residual O($a$) cutoff effects of several 
step-scaling functions and we compute renormalization
factors at the matching scale. As an example
of possible application of the $\chi$SF we compute the renormalized strange quark mass using large volume data
obtained from Wilson twisted mass fermions at maximal twist.

\end{abstract}


\end{frontmatter}
\cleardoublepage


\section{Introduction}
\label{sec:intro}

The Schr\"odinger functional (SF) scheme~\cite{Luscher:1992an,Sint:1993un,Luscher:2006df} 
has been widely employed for performing non-perturbative renormalization and scaling studies.
For an incomplete list of 
references see~\cite{Luscher:1993gh,Capitani:1998mq,Guagnelli:2003hw,Pena:2004gb,DellaMorte:2004bc,DellaMorte:2005kg,Aoki:2009tf,Aoki:2010wm,Tekin:2010mm}.

It is not straightforward to implement a Schr\"odinger functional (SF) scheme that
retains the property of automatic O($a$) improvement~\cite{Frezzotti:2003ni} for massless Wilson fermions.
A solution to this problem has been proposed recently by Sint~\cite{Sint:2010eh},
called the chirally rotated Schr\"odinger functional ($\chi$SF). 
In refs.~\cite{Lopez:2008ns,Lopez:2009yc,Sint:2010xy}
preliminary studies of the $\chi$SF were presented and, recently in a companion paper~\cite{Lopez:2012as},
we have performed a detailed study of the non-perturbative tuning of the $\chi$SF in a quenched setup. 
To retain the property of bulk automatic
O($a$) improvement, two parameters of the lattice action, one denoted by $\zf$ and the other,
the usual hopping parameter $\kappa$, have to be tuned non-perturbatively to their critical values.

In ref.~\cite{Lopez:2012as} we have defined a tuning procedure and we have 
found that the tuning of $\zf$ and $\kappa$ is numerically feasible
with standard techniques. 
We have found that several different tuning conditions lead to fully compatible results
for all physical correlation functions and we have also numerically checked that in the continuum limit
the correct boundary conditions, i.e. the boundary conditions needed to have bulk automatic
O($a$) improvement, are recovered.

The $\chi$SF, being compatible with automatic O($a$) improvement, can hence be used for renormalizing
bare operators computed with Wilson twisted mass fermions at maximal twist~\cite{Boucaud:2007uk}
and the computation of operator specific improvement coefficients can be avoided.
In this paper we want to demonstrate this explicitly for the particular case of a twist-2
operator. Additionally, we discuss the application of the $\chi$SF scheme
for the renormalization of quark masses.

The computation of such quantities allows us to perform
a test of the universality of the continuum limit of lattice QCD and,
as a result, to confirm the correctness of the continuum limit of the novel $\chi$SF scheme itself.
The latter can be achieved by comparing the continuum limit values of the 
computed quantities in the $\chi$SF formulation 
to those values obtained using the standard version of the SF.
Since, as discussed in ref.~\cite{Sint:2010eh,Lopez:2012as}, the SF and the 
$\chi$SF are the same formulations in the continuum,
the final results in the continuum limit, as obtained from the two formulations, should be exactly the same
given a common choice of a renormalization prescription.

The paper is organized as follows. We discuss the continuum limit of the step-scaling function (SSF)
of the pseudoscalar density, $\sigma_{\mathrm{P}}$, in sect.~\ref{sec:ssf} and 
of the non-singlet twist-2 operators, $\sigma_{\mathrm{O}_{12}}$ and $\sigma_{\mathrm{O}_{44}}$, in sect.~\ref{sec:twist2}.
The continuum limit values are then compared to the ones obtained using the SF,
with which, as expected, we find agreement.
Next, in sect.~\ref{sec:zfactors} we show the results of the determination of the renormalization factors
of the pseudoscalar density, $Z_{\mathrm{P}}$,
and the twist-2 operators, $Z_{\mathrm{O}_{12}}$ and $Z_{\mathrm{O}_{44}}$,
at hadronic matching scales and non-zero lattice spacing.
For the latter, results are compared against values obtained using the
standard formulation of the SF with two different regularizations,
non-improved Wilson and clover improved Wilson fermions.
Eventually in sect.~\ref{sec:ms} we compute the running strange quark mass,
using the values of $Z_{\mathrm{P}}$ obtained from the $\chi$SF scheme
and the tuned bare quark mass obtained using twisted mass Wilson fermions at maximal twist.
We compare our findings with the strange quark mass obtained using 
standard SF as a renormalization scheme and non-perturbatively improved Wilson fermions.
All these results are, therefore, a demonstration of the correctness of
the continuum limit of the $\chi$SF renormalization scheme and,
consequently, of its applicability in the determination of renormalization factors.

The analysis presented in this paper relies substantially on the
results obtained in ref.~\cite{Lopez:2012as}. We therefore assume from now on that the reader is familiar
with that paper and the notation adopted there will be taken over without further notice. 
Additionally all the equations of ref.~\cite{Lopez:2012as} will be denoted by the equation number 
prefixed by I as for example (I.5.10).

\section{Step-scaling functions: pseudoscalar density}
\label{sec:ssf}

The SSF, $\sigma_{\mathrm{O}}$, of a scale-dependent
observable, $O(L)$, describes the behavior of $O(L)$ under changes in
the value of the renormalization scale, $\mu=1/L$, where $L$ denotes the 
linear extent of the finite volume.
The reason SSFs are good candidates to perform universality tests is that
they are finite quantities that depend upon the
renormalization scheme and the renormalization prescription employed.
While at finite values of the cutoff, SSFs are regularization-dependent, 
after the removal of the cutoff they are independent of the
regulator.

As a simple example of a SSF, we consider the  
normalization $\zp$ of the pseudoscalar density.
The evolution of $\zp$ from a scale $L$ to $sL$
is described by the step-scaling function defined by
\be
\zp(L)\sip(s,\overline{g}^{2}(L)) =\zp(sL)\,,
\ee
where the $\overline{g}^{2}(L)$ is the renormalized coupling at the scale $L$.
Since the running of the coupling is needed
to compute the SSF of $Z_P$, or more generally any operator, then the
SSF of the gauge coupling itself is needed for the computation of
any SSF.
In this work, the quenched setup is chosen and the gauge action
is the Wilson gauge action, so there is no need to recompute 
the renormalized gauge coupling and
its corresponding SSF. These are already known from previous
publications~\cite{Luscher:1993gh,Capitani:1998mq,Guagnelli:2004za} 
for a very wide range of energies.
If a different gauge action or dynamical fermions
were included, the gauge coupling would need to be recomputed
with the new formulation.

The previous discussion refers to the formal continuum theory. Using a 
hypercubic lattice with spacing $a$,
the chosen renormalization condition for the pseudoscalar density in the
$\chi$SF scheme is
\begin{equation}
\label{eq:DefZPChiSF}
Z_{\textrm{P}}(g_{0},L/a) = c(\theta,a/L) \,
\frac{\sqrt{g_{1}(\theta)}}{g_{\textrm{P}}(L/2,\theta)}\Big\vert_{m=0}
\, .
\end{equation}
In this expression, $m=0$ indicates that the renormalization
condition is imposed at zero quark mass. In our case, corresponding to Wilson fermions,
this means at the critical value of the bare quark mass, $m_{0}=\mcr$, as
determined from the tuning in ref.~\cite{Lopez:2012as}.
The factor $c(\theta,a/L)$ is chosen such that
$Z_{\textrm{P}}$ takes the correct value at tree-level,
$Z_{\textrm{P}}(0,L/a)=1$. Therefore it is defined as
\begin{equation}
\label{eq:DefCZP}
c(\theta,a/L) \equiv
\frac{g_{\mathrm{P}}(L/2,\theta)}{\sqrt{g_{1}(\theta)}} \Big\vert_{m_{0}=0}^{\textrm{tree}}
\, .
\end{equation}
The two-point functions entering the definition of $Z_{\mathrm{P}}$
have already been discussed in ref.~\cite{Lopez:2012as}.
They are
\begin{equation}
\label{eq:gPZP}
g_{\textrm{P}}(x_{0},\theta) \equiv
g_{\textrm{P}_{-}}^{11}(x_{0},\theta) = - \frac{a^{3}}{L^{3}}\sum_{\bx}\langle
P^{1}(x)\widetilde{\mathcal{P}}_{-}^{1}\rangle\,,
\end{equation}
with $P^{1}$ denoting the pseudoscalar density
and
\begin{equation}
\label{eq:g1ZP}
g_{1}(\theta) \equiv g_{1}^{11}(\theta) = -\frac{1}{L^{6}}\, \langle
{\widetilde{\mathcal{P}'}}_{+}^{1}\widetilde{\mathcal{P}}_{-}^{1}
\rangle \, .
\end{equation}
See ref.~\cite{Lopez:2012as} for the remaining details of eqs.~(\ref{eq:gPZP},\ref{eq:g1ZP}).
In order to determine the renormalization prescription completely, a
value of $\boldsymbol{\theta}$ has to be chosen. In particular, we consider here
two cases, $\boldsymbol{\theta}=(0.5,0.5,0.5)$ and $\boldsymbol{\theta}=(1,0,0)$.

Given a fixed value of the renormalization scale, defined through
$\overline{g}^{2}(L)=u$, and a fixed value of the lattice spacing, leading hence 
to a fixed value of $L/a$, the lattice SSF of the pseudoscalar density is given by
\begin{equation}
\label{eq:SSFZPLat}
\Sigma_{\textrm{P}}(s,u,a/L) =
\frac{Z_{\textrm{P}}(g_{0},sL/a)}{Z_{\textrm{P}}(g_{0},L/a)} \Big
\vert_{m=0,\, \overline{g}^{2}(L)=u} \, .
\end{equation}
In the continuum limit, the SSF is finite and takes the value
\be
\label{eq:SSFZPCon}
\sigma_{\textrm{P}}(s,u) = \lim_{a \rightarrow 0} \Sigma_{\textrm{P}}(s,u,a/L)\,.
\ee

From the definition of the SSF, eq.~\eqref{eq:SSFZPLat},
it follows that in order to compute the SSF at a certain value
of the renormalization scale, $1/L$, the $Z$-factor needs to be
evaluated at both $L$ and $sL$ at the same value of the lattice
spacing. In this work we always use $s=2$.
This means that for a fixed value of $a$ (equivalently $\beta=6/g_0^2$),
simulations have to be performed at a certain value of $L/a$ and also at $2L/a$.
In such computations, the values of all parameters,
e.g. $\kcr$ and $z_{f}^{c}$, are the same at $L$ and $2L$ for
a fixed value of the lattice spacing since these parameters only
depend on the bare coupling.
This is important because it implies that the tuning of the parameters
only needs to be performed on the smaller lattices. 

We summarize the results for $Z_{\mathrm{P}}$ in
tab.~\ref{tab:ZP},
at $\boldsymbol{\theta}=(0.5,0.5,0.5)$ and $\boldsymbol{\theta}=(1,0,0)$.
Results at three different values of the renormalization scale are
presented; the hadronic scale $L=1.436\,r_{0}$, 
the intermediate scale $\overline{g}^{2}=2.4484$ and the
perturbative scale $\overline{g}^{2}=0.9944$.
The computation of the SSF is performed only at the intermediate
and the perturbative scales.
We use the results for $\zp$ at the hadronic (matching) scale
for the calculation of the renormalized strange quark mass as discussed in sect.~\ref{sec:ms}.
All results presented in the present and following sections have been
obtained using the critical values of the parameters, $\kcr$ and
$z_{f}^{c}$, as determined from the tuning condition called (1*) in ref.~\cite{Lopez:2012as}.

From the values in tab.~\ref{tab:ZP}, we have computed the SSF
for each lattice spacing at the intermediate and perturbative scales.
The results are presented in tab.~\ref{tab:LSSFTheta0.5} for
$\boldsymbol{\theta}=(0.5,0.5,0.5)$ and 
tab.~\ref{tab:LSSFTheta100} for $\boldsymbol{\theta}=(1,0,0)$.
In tab.~\ref{tab:LSSFTheta0.5} we also show the results obtained from
the SF with improved and standard Wilson fermions~\cite{Guagnelli:2004za}.

For both the $\chi$SF and the SF with clover improved Wilson fermions, we have performed the continuum limit 
with a linear fit of the SSF in $(a/L)^{2}$, i.e. using a form
\begin{equation}
\label{eq:Sigmafit_quad}
y =c_{0} + c_{1}\Big(\frac{a}{L}\Big)^{2}\,.
\end{equation}
While a linear fit in $a/L$ for the SF with standard Wilson fermions was used,
\begin{equation}
\label{eq:Sigmafit_lin}
\bar{y} =\bar{c}_{0} + \bar{c}_{1}\Big(\frac{a}{L}\Big) \, .
\end{equation}
The results of our fits are summarized in
tab.~\ref{tab:SSFcont.Theta0.5} for $\boldsymbol{\theta}=(0.5,0.5,0.5)$ 
and in tab.~\ref{tab:SSFcont.Theta100} for $\boldsymbol{\theta}=(1,0,0)$.
In both tables we show the results at the two values of the
renormalization scale that have been considered,
$\overline{g}^{2}=2.4484$ and $\overline{g}^{2}=0.9944$.
For comparison, in tab.~\ref{tab:SSFcont.Theta0.5} we also present the
continuum limit results for the SF with improved and standard Wilson
fermions. We have performed our own fits of the values obtained from the
SF, since in~\cite{Guagnelli:2004za} there are no tables with the
final continuum limit values, where we could read the data from.

\begin{figure}
\centering
\includegraphics[width=0.8\textwidth]{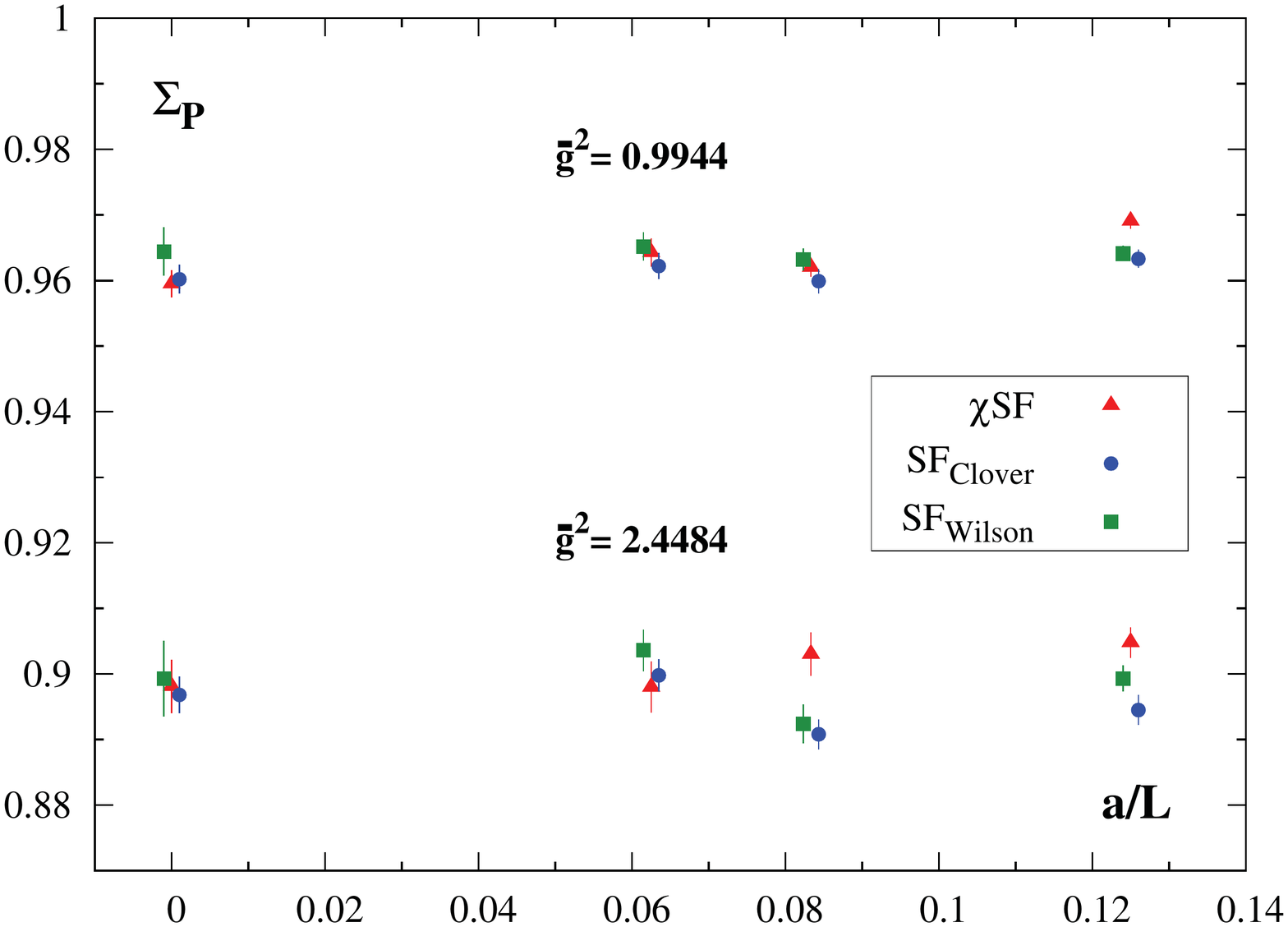}
\caption{Lattice SSF of the pseudoscalar density and continuum limit values.
  Results are shown for the $\chi$SF with standard Wilson fermions
  and the SF with improved and standard Wilson fermions,
  at the intermediate and perturbative scales
  and for $\boldsymbol{\theta}=(0.5,0.5,0.5)$.
  The extrapolations to the continuum limit are linear in $(a/L)^{2}$ for the
  $\chi$SF and the SF with improved fermions and linear in $a/L$ for the
  SF with standard Wilson fermions. 
  The numerical values are provided in tab.~\ref{tab:SSFcont.Theta0.5}.
  The values from the SF~\cite{Guagnelli:2004za} 
  have been slightly displaced to the right and
  left for the improved and the standard Wilson fermions formulations, respectively.}
\label{fig:SSF.Theta0.5}
\end{figure}

A comparison of the SSF for the different lattice fermions
is shown in fig.~\ref{fig:SSF.Theta0.5}.
There we plot the results for $\Sigma_{\mathrm{P}}$
(cf. tab.~\ref{tab:LSSFTheta0.5}) as a function of $a/L$ for the three
formulations.
The corresponding continuum limit values are also shown
(cf. tab.~\ref{tab:SSFcont.Theta0.5}).
We can see that the slopes, i.e. the values of $c_1$ and $\bar{c}_1$ in 
eqs.~(\ref{eq:Sigmafit_quad},\ref{eq:Sigmafit_lin}),
are consistent with
zero in each case.
Furthermore the results of the three regularizations agree in the
continuum limit at both values of the renormalization scale.
Moreover, at non-zero lattice spacing the results for the three formulations agree at $L/a=16$ for the
intermediate scale and at $L/a=12,16$ for the perturbative scale.

\begin{figure}
\centering
\includegraphics[width=0.8\textwidth]{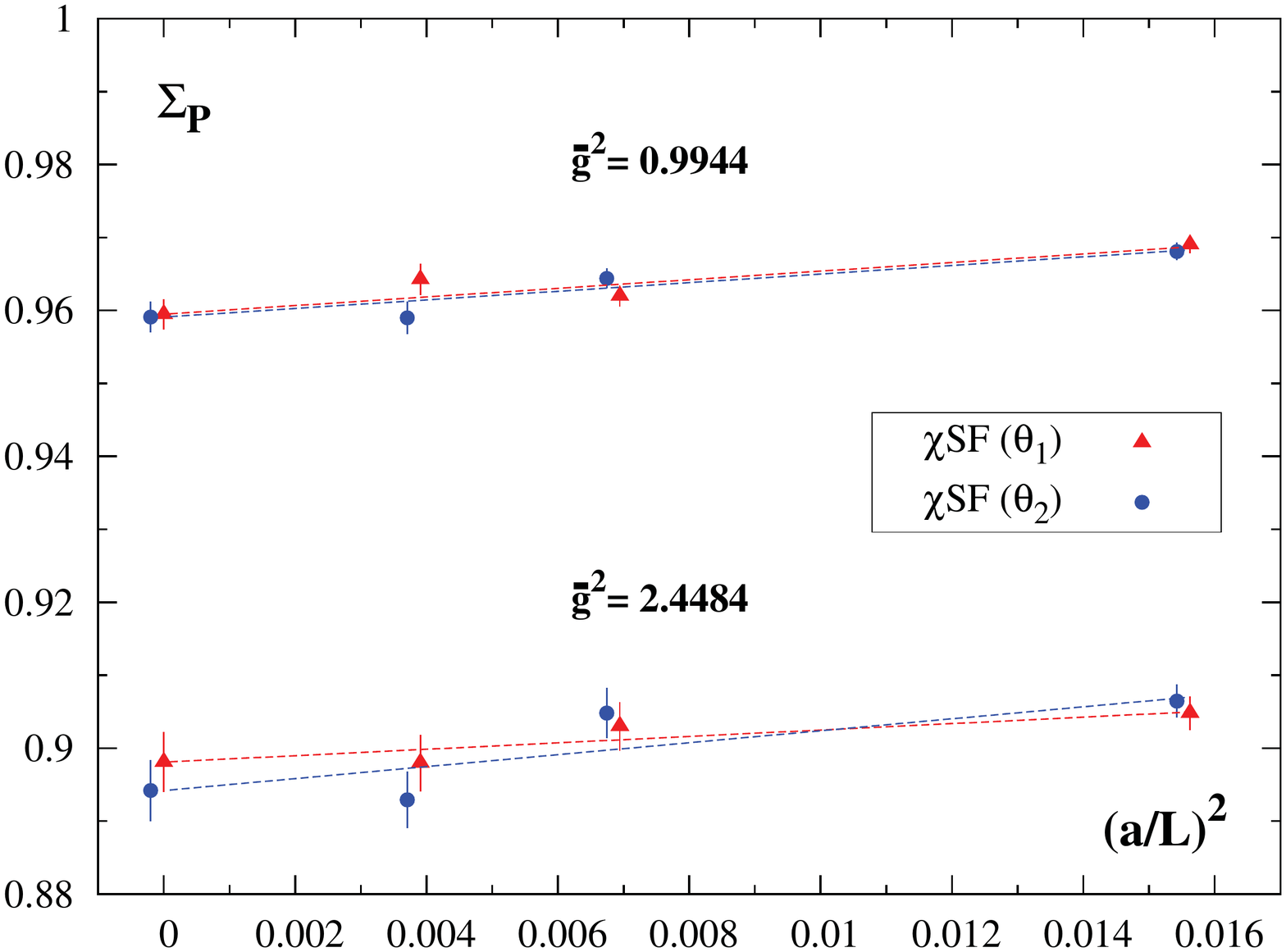}
\caption{ Continuum limit extrapolations of the SSF for the
  pseudoscalar density.
  The $\chi$SF results for both
  $\theta_{1}\equiv\boldsymbol{\theta}=(0.5,0.5,0.5)$
  and
  $\theta_{2}\equiv\boldsymbol{\theta}=(1,0,0)$ are shown  
  at the intermediate and perturbative scales.
  The extrapolations to the continuum limit are linear in $(a/L)^{2}$
  and plotted as dashed lines.
  The fit results are provided in tab.~\ref{tab:SSFcont.Theta0.5} 
  and tab.~\ref{tab:SSFcont.Theta100}.
  The values in the continuum limit are also plotted.
  The results from $\theta_{2}$ have been slightly displaced to the left.}
\label{fig:SSFall.ChiSF}
\end{figure}

In fig.~\ref{fig:SSFall.ChiSF} we show the extrapolation to the
continuum limit of $\Sigma_{\mathrm{P}}$ as determined from the
$\chi$SF, now for both values of $\boldsymbol{\theta}$.
The results are plotted as a function of $(a/L)^{2}$
and the corresponding values in the continuum limit,
$\sigma_{\mathrm{P}}$, are also shown.
The fitting curves are also plotted.
Note that for the
two values of $\boldsymbol{\theta}$ employed in the definition of the
renormalization prescription, the continuum limit values should not
be compared. Different values of $\boldsymbol{\theta}$ give rise to
different renormalization prescriptions. Consequently, the
results are not expected to agree in the continuum limit, even though the 
values appear quite similar.

\begin{figure}
\centering
\includegraphics[width=0.8\textwidth]{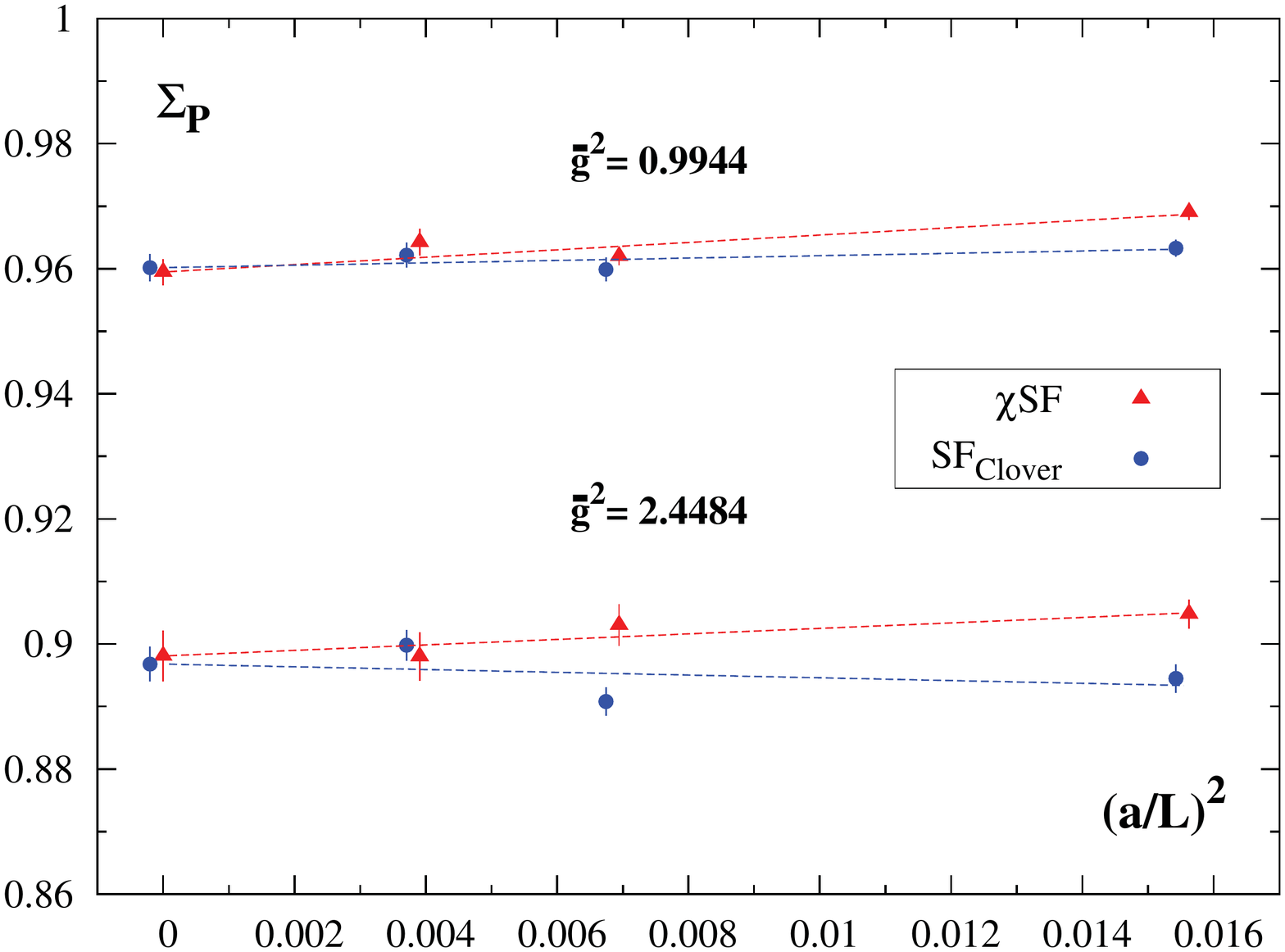}
\caption{Continuum limit extrapolation of the SSF of the pseudoscalar
  density.
  Results are shown for the $\chi$SF with standard Wilson fermions
  and the SF with improved Wilson fermions,
  at the intermediate and perturbative scales
  and for $\boldsymbol{\theta}=(0.5,0.5,0.5)$.
  The extrapolations to the continuum limit, plotted as dashed lines, are linear in $(a/L)^{2}$
  for both cases. The results from the fits are presented in
  tab.~\ref{tab:SSFcont.Theta0.5}.
  The values from the SF have been slightly displaced to the left.}
\label{fig:SSFimpr.Theta0.5}
\end{figure}

A similar plot is shown in fig.~\ref{fig:SSFimpr.Theta0.5},
where we compare the results of the extrapolation to the continuum
limit for the $\chi$SF and the improved SF.
Results are shown at the two values of the renormalization scale and
for $\boldsymbol{\theta}=(0.5,0.5,0.5)$.
Since the same renormalization prescription is chosen in both
formulations, the continuum limit values should now agree
between the SF and the $\chi$SF.
As seen in the plot, this is indeed the case.
Note that the slopes for the two formulations have similar absolute
values but opposite signs.
This suggests that a constrained continuum limit extrapolation 
would reduce the uncertainty of the final results.

From the results presented here, we conclude that the continuum
limit of the SSF of the pseudoscalar density determined from the $\chi$SF
agrees with that obtained from
the standard formulation of the SF, both with and without
improvement.
These results are therefore a successful test of the universality of
the continuum limit.
Moreover, the approach to the continuum limit is consistent with the expected
leading $O(a^{2})$ discretization effects in the $\chi$SF formulation, 
indicating that the $\chi$SF is indeed compatible with bulk automatic
$O(a)$-improvement.

\section{Step-scaling functions: twist-2 operators}
\label{sec:twist2}

The previous discussion for the pseudoscalar density
can be repeated for any other scale-dependent observable.
In particular, we now compute the 
SSF of two lattice
realizations of the non-singlet twist-two operator whose hadronic
matrix elements are related to the lowest non-trivial moment of the
corresponding unpolarized structure function.

In Minkowski space the relevant twist-2 gauge-invariant composite operator is
\begin{equation}
\label{eq:Twist2OpDefMink}
\mathrm{i}^{n-1} \, \overline{\psi}(x)\gamma_{\left\{\mu_{1}\right.}\lrD_{\mu_{2}}\cdots \lrD_{\left.\mu_{n}\right\}}\, \frac{\tau^{a}}{2}\psi(x) +
\textrm{`trace terms'} \, .
\end{equation}
The symmetrization in the Lorentz indices, $\left\{ \ldots \right\}$,
is required because we deal here with unpolarized
scattering and the `trace terms' (terms with $g_{\mu_{i}\mu_{j}}$) are needed
in order to provide the composite field with a definite spin.
The covariant derivative, $\lrD_{\mu}$, is defined as the combination
\begin{equation}
\label{eq:LRSymDer}
\lrD_{\mu} = \frac{\overrightarrow{D}_{\mu} - \overleftarrow{D}_{\mu}}{2}
\, ,
\end{equation}
with $\overrightarrow{D}_{\mu}$ and $\overleftarrow{D}_{\mu}$ the
covariant derivatives acting to the right and left, respectively.

Since we compute quantities within the $\chi$SF scheme and we work in
the twisted basis, we give here the explicit expressions of the twist-2
operators in Euclidean space and in the twisted basis.
These can be obtained from the
corresponding expressions in the physical basis by
applying the standard axial rotation eq.~(I.2.2)
in the continuum theory and then directly translating the fields and
derivatives to the lattice.
This results in
\begin{equation}
\label{eq:Twist2OpFirstMomLattTwFinal}
O_{\mu\nu}^{a}(x)
= \overline{\chi}(x) \, e^{\mathrm{i}\frac{\alpha}{2}\gamma_{5}\tau^{3}}\,
 \gamma_{\left\{\mu \right.} \lrD_{\left. \nu \right\}} \, \frac{\tau^{a}}{2}\,
e^{\mathrm{i}\frac{\alpha}{2}\gamma_{5}\tau^{3}}\, \chi(x)\,.
\end{equation}
Depending on the flavor structure,
\begin{alignat}{2}
  \label{eq:Twist2OpFirstMomLattTwFlavorFinal}
  O_{\mu\nu}^{a}(x) = \, 
    \begin{cases}
      \overline{\chi}(x) \,   \gamma_{\left\{\mu \right.}
      \lrD_{\left. \nu \right\}} \,
      \Big[ \cos(\alpha)\, \frac{\tau^{a}}{2}
      + \epsilon_{ab3}\sin(\alpha)\, \gamma_{5} \, \frac{\tau^{b}}{2}
      \Big] \chi(x)
      & (a=1,2) \, ,\\
      \\
      \overline{\chi}(x) \, \gamma_{\left\{\mu \right.}
      \lrD_{\left. \nu \right\}} \,\frac{\tau^{a}}{2}\, \chi(x)
      & (a=3) \, ,
    \end{cases}
  \end{alignat}
with $\epsilon_{abc}$ the totally anti-symmetric tensor ($\epsilon_{123}=1$).
In the particular case of maximal twist, $\alpha=\pi/2$, this
expression reduces to
\begin{alignat}{2}
  \label{eq:Twist2OpFirstMomLattTwMaxFlavorFinal}
  O_{\mu\nu}^{a}(x) = \,
    \begin{cases}
      \epsilon_{ab3}\, \overline{\chi}(x) \,   \gamma_{\left\{\mu
  \right.} \lrD_{\left. \nu \right\}} \gamma_{5} \, \frac{\tau^{b}}{2}\,
\chi(x)
& (a=1,2) \, ,\\
\\
\overline{\chi}(x) \, \gamma_{\left\{\mu \right.}
\lrD_{\left. \nu \right\}} \,\frac{\tau^{a}}{2}\, \chi(x)
& (a=3) \, .
\end{cases}
\end{alignat}

These composite fields are scale-dependent quantities that need to be
renormalized, $O_{\mathrm{R}}=Z_{\mathrm{O}}^{-1}\,O_{\mathrm{B}}$,
where $O_{\mathrm{B}}$ is either bare operator from eq.~\eqref{eq:Twist2OpFirstMomLattTwMaxFlavorFinal} and $Z_{\mathrm{O}}$ is the corresponding renormalization constant.

The corresponding SSF in the continuum $\sigma_{\mathrm{Z}_\mathrm{O}}(s,\overline{g}^{2}(L))$ is defined as
\be
\label{eq:RGEOintegrated}
Z_{\mathrm{O}} (sL) = \sigma_{\mathrm{Z}_\mathrm{O}}(s,\overline{g}^{2}(L)) \, Z_{\mathrm{O}}(L)\,.
\ee
Using a lattice regulator,
lattice artifacts must be taken into account and the
renormalized operator is given as\footnote{The definition of the renormalized operator with $Z^{-1}$ is done to be consistent with
the definitions used in ref.~\cite{Guagnelli:2003hw}.}
\begin{equation}
\label{eq:RenOLat}
O_{\mathrm{R}}(L) = \lim_{a \rightarrow 0}
Z_{\mathrm{O}}^{-1}(g_{0},L/a) \, O_{\mathrm{B}}(g_{0}) \, .
\end{equation}

Eventually, we will compute boundary to bulk correlation functions
with the twist-2 operators inserted in the bulk of the lattice at
some space-time point $x$.
The boundary interpolating fields at $x_{0}=0$ that we consider here
are, following~\cite{Bucarelli:1998mu}, expressed in the physical basis,
\begin{equation}
\label{eq:bopXOmunuPhys}
a^{6}\sum_{\by,\bz} 
\overline{\zeta}(\by)\gamma_{k}\frac{\tau^{a}}{2}\zeta(\bz)\, .
\end{equation}
Performing a rotation to the twisted basis, with maximal twist angle,
such boundary interpolating fields take the form
\begin{alignat}{2}
\label{eq:bopXOmunuTwMax}
\mathcal{O}_{\gamma_{k}}^{a}= \,
\begin{cases}
\epsilon_{ab3} \, a^{6}\, \sum_{\by,\bz}\,
\overline{\zeta}(\by)\, \gamma_{k}\gamma_{5}\frac{\tau^{b}}{2}\, \widetilde{Q}_{-}\,\zeta(\bz)
& (a=1,2) \, ,\\
\\
a^{6}\, \sum_{\by,\bz}\, \overline{\zeta}(\by)
\, \gamma_{k}\frac{\tau^{a}}{2}\, \widetilde{Q}_{-}\,\zeta(\bz)
& (a=3) \, .
\end{cases}
\end{alignat}
In particular, we consider two cases for the gamma matrices at the
boundaries, $\gamma_{k}$ with $k=1,2$. The case $k=1$ ($k=2$) will be
used when computing the correlation function of the operator $O_{44}^{a}$
($O_{12}^{a}$).
Finally, the correlation functions that we consider here are the
following,
\begin{subequations}
\label{eq:gOmunu}
\begin{align}
g_{12}(x_{0}, \theta) & \equiv
- \frac{a^{9}}{L^{3}}\, \sum_{\bx,\by,\bz}\, \langle
\overline{\chi}(x) \,   \gamma_{\left\{1 \right.} \lrD_{\left. 2 \right\}} \gamma_{5} \, \frac{\tau^{1}}{2}\,\chi(x)\,
\overline{\zeta}(\by)\, \gamma_{2}\gamma_{5}\frac{\tau^{1}}{2}\, \widetilde{Q}_{-}\,\zeta(\bz)
\rangle \label{eq:gO12} \, , \\
g_{44}(x_{0}, \theta) & \equiv
- \frac{a^{9}}{L^{3}}\, \sum_{\bx,\by,\bz}\, \langle
\overline{\chi}(x) \,   \gamma_{\left\{0 \right.} \lrD_{\left. 0 \right\}} \gamma_{5} \, \frac{\tau^{1}}{2}\,\chi(x)\,
\overline{\zeta}(\by)\, \gamma_{1}\gamma_{5}\frac{\tau^{1}}{2}\, \widetilde{Q}_{-}\,\zeta(\bz)
\rangle \label{eq:gO44} \, .
\end{align}
\end{subequations}
Note that we have chosen only the cases where the two flavor matrices,
in the bulk and at the boundary,
are the same and we have picked up only the component $\tau^{1}$.
The reason for choosing both matrices to be the same is that, 
due to symmetries, this is the only combination for which the
correlation function does not vanish.
Amongst the three possibilites, $\tau^{1,2,3}$,
each of them should lead to the same value in the continuum limit.
However, due to our particular setup where flavor
symmetry is broken at finite lattice spacing, there is a distinction between $\tau^{1,2}$ and
$\tau^{3}$.
Choosing $\tau^3$ would lead to simpler looking expressions, but the appearance of computationally
demanding disconnected diagrams leads us to opt for $\tau^{1,2}$.
Since these two cases are exactly equivalent, we can select just $\tau^{1}$.

We now specify a renormalization prescription for the twist-2
operator within the $\chi$SF scheme.
In particular we impose the renormalization condition
\begin{equation}
\label{eq:DefZOChiSF}
Z_{\mathrm{O}}(g_{0},L/a) = c(\theta,a/L) \,
\frac{g_{\mathrm{O}}(L/2,\theta)}{\sqrt{g_{1}(\theta)}}
\Big\vert_{m=0} \, .
\end{equation}
The factor $c(\theta,a/L)$ is chosen such that
$Z_{\textrm{O}}$ takes the correct value at tree-level, $Z_{\textrm{O}}(0,L/a)=1$. Therefore it is defined as
\begin{equation}
\label{eq:DefCZO}
c(\theta,a/L) \equiv
\frac {\sqrt{g_{1}(\theta)}}{g_{\mathrm{O}}(L/2,\theta)}
\Big\vert_{m_{0}=0}^{\textrm{tree}} \, .
\end{equation}
In this expression, $g_{1}$ is the two-point function defined in eq.~\eqref{eq:g1ZP}.
The other two-point function, $g_{\mathrm{O}}$, is either
$g_{12}$ or $g_{44}$ from eq.~\eqref{eq:gOmunu},
corresponding to either $O_{12}^{a}$ or $O_{44}^{a}$, respectively.
In order to fix the renormalization prescription completely, a
value of $\boldsymbol{\theta}$ has to be chosen. In particular, we consider again the
two cases $\boldsymbol{\theta}=(0.5,0.5,0.5)$ and $\boldsymbol{\theta}=(1,0,0)$.
The reason for studying the case with $\boldsymbol{\theta}=(1,0,0)$ is that
this is the only choice with  $\boldsymbol{\theta}\ne \boldsymbol{0}$
for which calculations with the standard SF have been performed~\cite{Guagnelli:2003hw}. 
This allows us to compare the continuum limit of the $\chi$SF.
Although there are no SF computations available for the choice
$\boldsymbol{\theta}=(0.5,0.5,0.5)$, we have also 
analyzed this setup since this value of theta is the usual choice for
calculations with the SF formulation.
Moreover, this additional choice for the parameter $\boldsymbol{\theta}$
allows us to examine the relative statistical uncertainties
when changing the renormalization prescription through $\boldsymbol{\theta}$.
All correlation functions are evaluated at $x_{0}=T/2$, where $T=L$ is
the time extent of the lattice.
The scale factor is always set to $s=2$.

Given a fixed value of the renormalization scale, defined through
$\overline{g}^{2}(L)=u$, and a fixed value of the lattice spacing $a$, leading to a fixed value
of $L/a$,
the lattice SSF of $O$, defined in the chiral limit, is given as
\begin{equation}
\label{eq:SSFZOLat}
\Sigma_{\mathrm{Z}_\mathrm{O}}(s,u,a/L) =
\frac{Z_{\mathrm{O}}(g_{0},sL/a)}{Z_{\mathrm{O}}(g_{0},L/a)} \Big
\vert_{m=0,\, \overline{g}^{2}(L)=u} \, .
\end{equation}
In the continuum limit the SSF is finite and takes the value
\begin{equation}
\label{eq:SSFZOCon}
\sigma_{\mathrm{Z}_\mathrm{O}}(s,u) = \lim_{a \rightarrow 0} \Sigma_{\mathrm{Z}_\mathrm{O}}(s,u,a/L) =
\frac{O_{\mathrm{R}}(L)}{O_{\mathrm{R}} (sL)} \Big \vert_{\overline{g}^{2}(L)=u} \, ,
\end{equation}
with $O_{\mathrm{R}}(L)$ the renormalized operator at a given value of
the physical scale $1/L$ computed in the $\chi$SF renormalization scheme.

We employ the definitions given above and the chosen renormalization prescription to determine the
renormalization factors of the operators $O_{12}^{a}$ and $O_{44}^{a}$
within the $\chi$SF scheme at finite lattice spacing.
Results are presented for several $\beta$-values and at three values
of the renormalization scale, $1/L$.
Two values of $L$ correspond to $\overline{g}^{2}=0.9944$ and
$\overline{g}^{2}=2.4484$ and the third is $L=1.436\,r_{0}$.
These results are given in tab.~\ref{tab:ZO12} and tab.~\ref{tab:ZO44}
for $O_{12}^{a}$ and $O_{44}^{a}$, respectively.
In both cases we provide results at both
$\boldsymbol{\theta}=(0.5,0.5,0.5)$ and $\boldsymbol{\theta}=(1,0,0)$.


\begin{figure}
\centering
\includegraphics[width=0.8\textwidth]{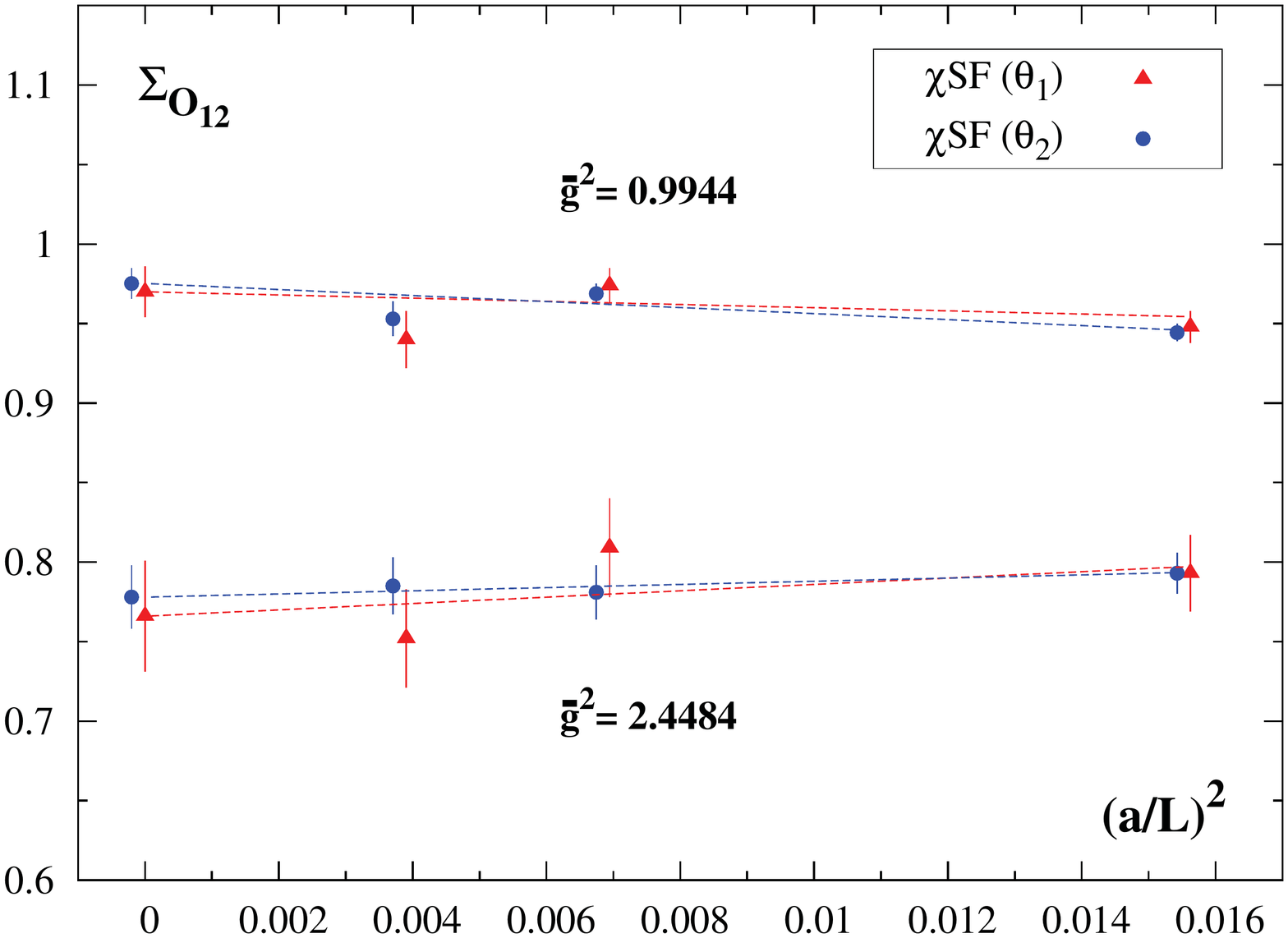}
\caption{Continuum limit extrapolation of the SSF of the operator $O_{12}$.
  Only $\chi$SF results are shown, both
  for
  $\theta_{1}\equiv\boldsymbol{\theta}=(0.5,0.5,0.5)$
 and
 $\theta_{2}\equiv\boldsymbol{\theta}=(1,0,0)$ and at both
  the intermediate and perturbative scales.
  The extrapolations to the continuum limit, plotted by dashed lines, 
  are linear in $(a/L)^{2}$. The fit results are given
  in tabs.~\ref{tab:SSFcont.O12O44Theta0.5}
  and~\ref{tab:SSFcont.O12O44Theta100}, and
  the values in the continuum limit are also plotted.
  The results for $\theta_{2}$ have been plotted slightly displaced to the left.}
\label{fig:SSFall_chiSF_O12}
\end{figure}

\begin{figure}
\centering
\includegraphics[width=0.8\textwidth]{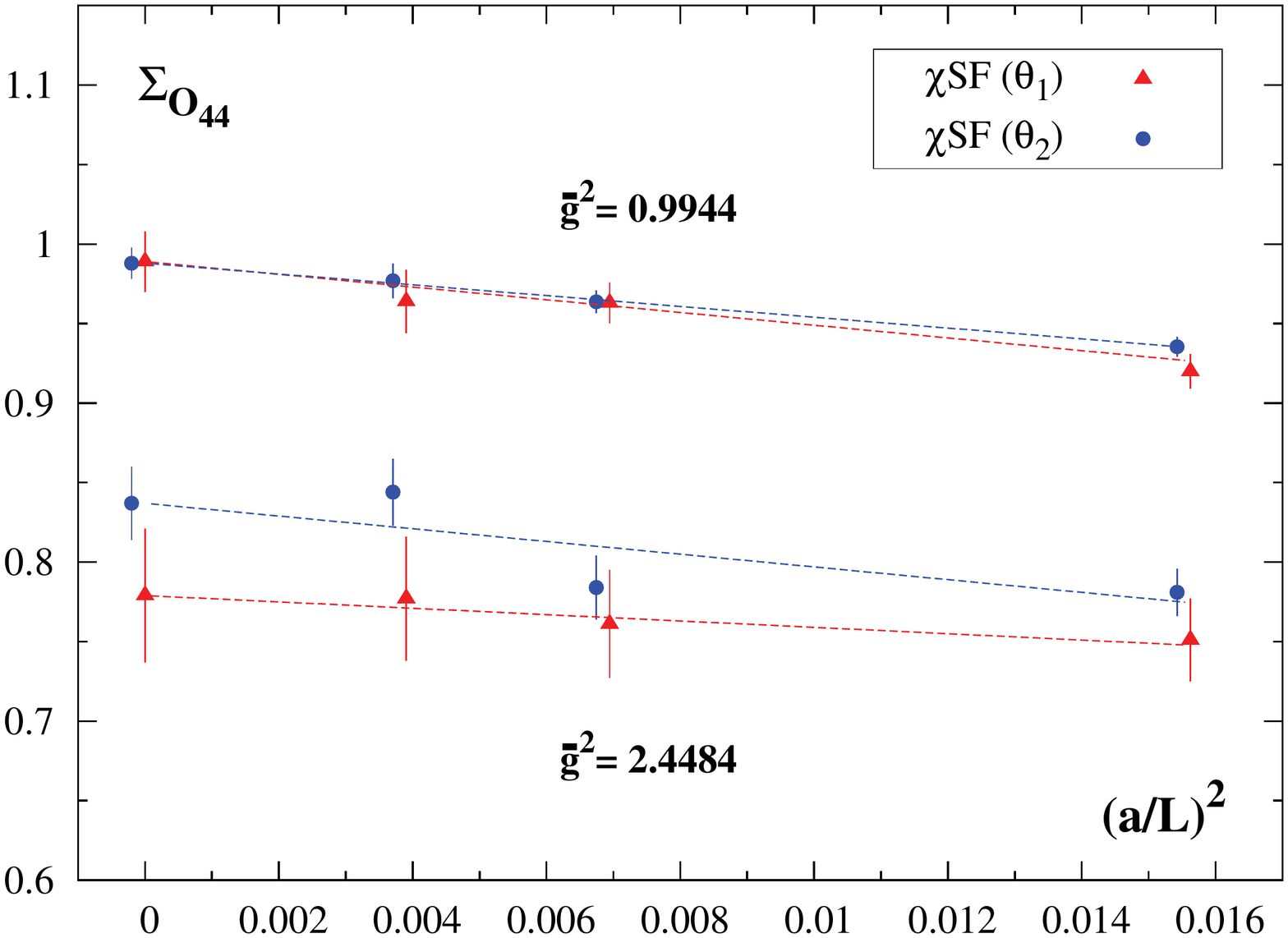}
\caption{Continuum limit extrapolation of the SSF of the operator $O_{44}$.
This is the same as fig.~\ref{fig:SSFall_chiSF_O12} but now for $O_{44}$ instead of $O_{12}$.}
\label{fig:SSFall_chiSF_O44}
\end{figure}

\begin{figure}
\centering
\includegraphics[width=0.8\textwidth]{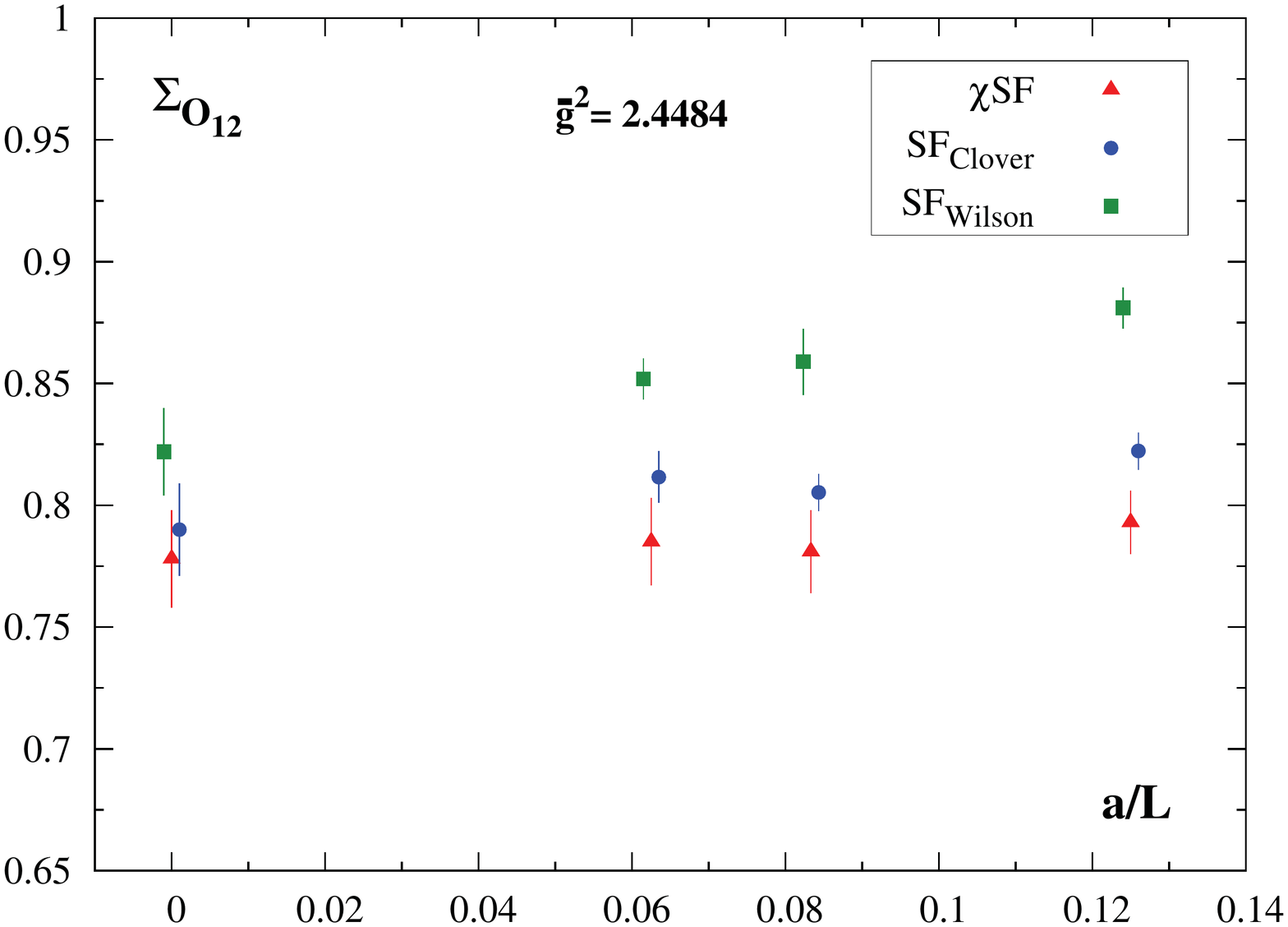}
\caption{Continuum limit approach of the SSF of the operator $O_{12}$.
  Results are shown for the $\chi$SF with standard Wilson fermions
  and for the SF with improved and standard Wilson fermions.
  The coupling corresponds to the intermediate scale and $\boldsymbol{\theta}=(1,0,0)$.
  The continuum limit is taken linear in $(a/L)^2$
  for the $\chi$SF and linear in $a/L$ for the SF with Wilson and clover improved Wilson 
  regularizations.  The fit results
  are provided in tab.~\ref{tab:SSFcont.O12O44Theta100} and the continuum limit values are also
  plotted.
  The data from the SF have been plotted slightly displaced to the right and
  left, respectively, for the improved and unimproved formulations.}
\label{fig:SSFimprO12}
\end{figure}

\begin{figure}
\centering
\includegraphics[width=0.8\textwidth]{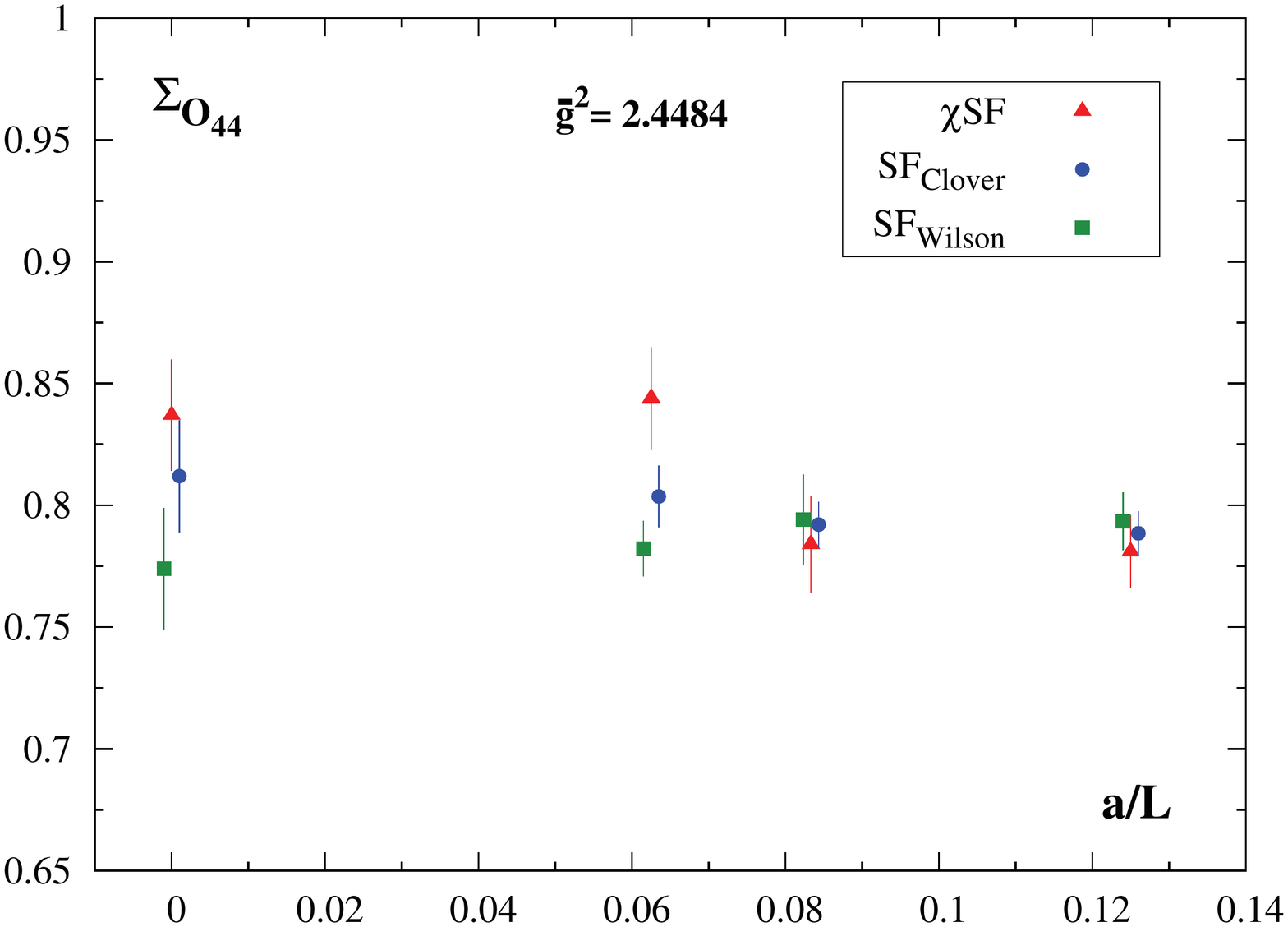}
\caption{Continuum limit approach of the SSF of the operator $O_{44}$.
  This is the same as fig.~\ref{fig:SSFimprO12} but now for $O_{44}$ instead of $O_{12}$. }
\label{fig:SSFimprO44}
\end{figure}

Concerning our results for the Z-factors in the $\chi$SF scheme,
tab.~\ref{tab:ZO12} and tab.~\ref{tab:ZO44},
at the three values of the renormalization scale and for all values of
the lattice spacing that we have analyzed, we now make several observations.
For either operator $O_{12}^{a}$ or $O_{44}^{a}$,
the relative statistical errors in the renormalization constants,
$\Delta Z_{\mathrm{O}}/Z_{\mathrm{O}}$,
are always smaller for $\boldsymbol{\theta}=(1,0,0)$ than for
$\boldsymbol{\theta}=(0.5,0.5,0.5)$ by nearly a factor of 2.
Moreover, at fixed values of all parameters,
the relative errors in $Z_{\mathrm{O}_{44}}$ are always slightly
larger than those of $Z_{\mathrm{O}_{12}}$.
These results are consistent with the pattern discussed 
previously in~\cite{Guagnelli:2003hw} within the standard SF setup,
where it was shown that for $|\boldsymbol{\theta}|\lesssim 1 $ the relative statistical errors
in $Z_{\mathrm{O}_{12}}$ and $Z_{\mathrm{O}_{44}}$ increase with decreasing
$|\boldsymbol{\theta}|$.
There it was also shown that the statistical errors in
$Z_{\mathrm{O}_{44}}$ are slightly larger than those in
$Z_{\mathrm{O}_{12}}$, which is consistent with what we observe from
our $\chi$SF calculation.

At the two most perturbative couplings,
$\overline{g}^{2}=0.9944$ and $\overline{g}^{2}=2.4484$,
we have determined the lattice SSFs for both operators.
The values are provided in
tab.~\ref{tab:LSSFO12O44Theta0.5} for $\boldsymbol{\theta}=(0.5,0.5,0.5)$ and
in tab.~\ref{tab:LSSFO12O44Theta100} for $\boldsymbol{\theta}=(1,0,0)$.
In tab.~\ref{tab:LSSFO12O44Theta100}, we have also added
the values obtained for the lattice SSFs using the SF scheme with
standard and non-perturbatively improved Wilson fermions.
These values were taken from~\cite{Guagnelli:2003hw}, where results are available
only at the intermediate coupling, $\overline{g}^{2}=2.4484$.

To take the continuum limit we have performed fits to our results in
tab.~\ref{tab:LSSFO12O44Theta0.5} and tab.~\ref{tab:LSSFO12O44Theta100},
linear in $(a/L)^{2}$ for the $\chi$SF formulation and linear in $a/L$
for the SF with both standard and improved Wilson fermions.
The fits for the SF calculation are linear in $a/L$ even for the formulation with improved
fermions because, although the action is improved in this
setup, the twist-2 operators themselves are not.
This is actually the major advantage of the $\chi$SF as a non-perturbative renormalization scheme: 
it is not necessary in this
setup to determine additional counterterms for the operators, since this formulation preserves
bulk automatic O($a$) improvement, up to boundary effects that are expected to be small. 
The results of these fits are presented in tab.~\ref{tab:SSFcont.O12O44Theta0.5}
and tab.~\ref{tab:SSFcont.O12O44Theta100}.

The results for the lattice SSFs from tabs.~\ref{tab:LSSFO12O44Theta0.5}
and~\ref{tab:LSSFO12O44Theta100}
are plotted in figs.~\ref{fig:SSFall_chiSF_O12}-\ref{fig:SSFimprO44},
where we show the continuum limit approach of all SSFs that we have
computed. In these figures we have also plotted the corresponding
values of the SSFs in the continuum limit and the fitting curves for the $\chi$SF case,
as given in tabs.~\ref{tab:SSFcont.O12O44Theta0.5}
and~\ref{tab:SSFcont.O12O44Theta100}.

In fig.~\ref{fig:SSFall_chiSF_O12}
and fig.~\ref{fig:SSFall_chiSF_O44}
we show the continuum limit approach of the SSFs of the operators
$O_{12}^{a}$ and $O_{44}^{a}$, respectively, within the $\chi$SF
scheme. In each figure, we plot the results for both values of
$\boldsymbol{\theta}$ and the two scales where we have computed the SSFs.
Results are presented as a function of $(a/L)^{2}$, since only
the $\chi$SF computation is considered.
We conclude, from these figures and the corresponding tables, that
the cutoff effects in the $\chi$SF SSFs are consistent with $O(a^{2})$
and are, in fact, small. We can also see that the discretization
effects are similar for different values of the renormalized coupling
and $\boldsymbol{\theta}$.
Note that the values in the continuum limit for different values of
$\boldsymbol{\theta}$ are not required to agree, since different
$\boldsymbol{\theta}$ values correspond to different renormalization
prescriptions.

In fig.~\ref{fig:SSFimprO12}
and fig.~\ref{fig:SSFimprO44}
we compare the results for the SSFs of $O_{12}^{a}$ and $O_{44}^{a}$
obtained from the three formulations, $\chi$SF and SF with
standard and improved Wilson fermions.
Only the computations for the intermediate coupling and for
$\boldsymbol{\theta}=(1,0,0)$ are plotted.
For a comparison with the SF,
the results are plotted as a function of $a/L$, although the values
in the continuum limit for the $\chi$SF have been obtained from
linear fits in $(a/L)^{2}$.
As explained earlier, we have performed the continuum
extrapolation of the SF results in~\cite{Guagnelli:2003hw}.
The results of each of these extrapolations
can be found in tab.~\ref{tab:SSFcont.O12O44Theta100}.
Additionally, the continuum limit extrapolations for $\boldsymbol{\theta}=(0.5,0.5,0.5)$
are provided in tab.~\ref{tab:SSFcont.O12O44Theta0.5} for the
$\chi$SF formulation.
There is good agreement, within the statistical uncertainties,
between the $\chi$SF and the improved SF formulations in the continuum
limit.

In summary, we conclude that there is agreement,
within the statistical errors,
in the continuum limit amongst the results from the three
formulations. 
This is another check of the universality of the continuum limit, this time
through the SSFs of a twist-2 operator.
Additionally, we observe that the scaling behavior of the
SSFs obtained from the $\chi$SF is consistent with leading $O(a^{2})$
discretization effects, which turn out to be rather small.

\section{Renormalization factors at hadronic scales}
\label{sec:zfactors}

Having established that the $\chi$SF does indeed maintain automatic bulk O($a$) 
improvement for several SSFs, we now discuss a physical application of these calculations.
In this section, we determine the renormalization factor at the hadronic
(matching) scale for the quark mass and for the twist-2 operators discussed in the
previous section. Then in the next section, 
we apply this to the determination of the strange quark mass.

With Wilson twisted mass (Wtm) fermions, the bare vector Ward identity~\cite{Frezzotti:2000nk} 
is exactly satisfied if one uses a point-split vector current. 
This implies that $Z_\mu = Z_P^{-1}$, where $Z_\mu$ is the bare quark mass renormalization and $Z_P$ is 
the renormalization constant of the
pseudoscalar density defined in eq.~\eqref{eq:DefZPChiSF}.
In the following, we denote the RGI quark mass by $M$, the bare
quark mass is $\mu_q(g_{0})$ and $\overline{\mu}_q(L)$ is the renormalized running quark mass
at the value $1/L$ of the renormalization scale.
The renormalized quark mass is given by
\be
\label{eq:Renmu}
\overline{\mu}_q(L) = Z_{\mathrm{P}}^{-1}(g_{0},L)\mu_q(g_{0}) \,.
\ee
The RGI quark mass can be related directly to the bare quark
mass by
\begin{equation}
\label{eq:Mofmbare}
M = Z_{\mathrm{M}}(g_{0})\mu_q(g_{0}) \,.
\end{equation}
The renormalization factor $Z_{\mathrm{M}}(g_{0})$ is defined as
the product of two terms as follows 
\begin{equation}
\label{eq:DefZM}
Z_{\mathrm{M}}(g_{0}) = \Bigg( \frac{M}{\overline{\mu}_q(L)}\Bigg) \Bigg(
\frac{1}{Z_{\mathrm{P}}(g_{0},L)} \Bigg)\, .
\end{equation}
The first term, $M/\overline{\mu}_q(L)$, is regularization independent but
depends on the renormalization scheme as well as on the chosen matching scale
$1/L$.
The second term, $Z_{\mathrm{P}}^{-1}(g_{0},L)$, depends on both the
renormalization scheme and the regulator.
The dependence is such that the RGI Z-factor $Z_{\mathrm{M}}(g_{0})$
does not depend on the renormalization scheme but only on the
regularization. All dependence on the matching scale has also
disappeared.

In the discussion above, all equations correspond to the continuum
theory. When the lattice is used as a regularization scheme, the correct
relation is
\begin{equation}
\label{eq:MofmbareLat}
M = Z_{\mathrm{M}}(g_{0})\mu_q(g_{0}) + O(a^{n}) \, ,
\end{equation}
with $n=1$ in case of unimproved formulations and $n=2$ if improvement
is applied.

The regularization independent part of $Z_{\mathrm{M}}(g_{0})$,
$M/\overline{\mu}_q(L)$,
has already
been determined in~\cite{Capitani:1998mq}.
The value is known in the continuum theory and at the matching
scale $L=1.436\,r_{0}$.
Once the continuum limit is performed, this factor is then universal,
i.e. regulator independent and we can use it directly in
our calculations without the need of a new computation since both the
SF and the $\chi$SF are equivalent formulations in the continuum theory.
The value obtained in~\cite{Capitani:1998mq} is
\begin{equation}
\label{eq:RegIndepPartZM}
M/\overline{\mu}_q(L) = 1.157\,(15) \quad \textrm{at} \quad L = 1.436\,r_{0} \, 
\end{equation}
with a relative error of $1.3\%$, which is sufficient for our purposes.

This means that we are left with the computation of two
quantities. One is the regularization dependent part of the total
renormalization factor, $Z_{\mathrm{P}}^{-1}(g_{0},L)$,
which must be computed at the matching scale $L$ of eq.~\eqref{eq:RegIndepPartZM}
for several values of $\beta$ and within the $\chi$SF scheme.
The other quantity is the bare quark mass, $\mu_q(g_{0})$,
which also has to be determined for a range of bare couplings
and is discussed shortly in
sect.~\ref{sec:ms}.

The determination of $Z_{\mathrm{P}}(g_{0},L/a)$ from the $\chi$SF, at a
certain value of the renormalization scale and for several values of
the lattice spacing, has already been explained in sect.~\ref{sec:ssf}
and the results are given in tab.~\ref{tab:ZP}.
Amongst the cases presented in tab.~\ref{tab:ZP}, we can restrict our
focus to the results corresponding to the matching scale, 
$L = 1.436\,r_{0}$ and $\boldsymbol{\theta}=(0.5,0.5,0.5)$.
For this choice of the parameters, we have computed
$Z_{\mathrm{P}}(g_{0},L/a)$ at several values of the lattice spacing
in the range $6.0 \le \beta \le 6.5$, which we recall in
tab.~\ref{tab:ZPmatching.Theta0.5}.

With these results, we determine a
smooth parameterization of the dependence of $Z_{\mathrm{P}}$ on $\beta$ for the range
covered in our calculation.
We use a polynomial fit to parameterize our results,
\begin{equation}
\label{eq:ZPmatchingbeta}
\begin{split}
&Z_{\mathrm{P}}(g_{0},L/a)_{L=1.436\, r_{0}} =
\sum_{i=0}^{2} z_{i}^{\textrm{P}}(\beta-6.0)^{i} \, ,\\
&\beta = 6/g_{0}^{2}, \qquad 6.0 \le \beta \le 6.5 \,.
\end{split}
\end{equation}
The fitted coefficients are provided in
tab.~\ref{tab:ZPbetaNP}.

This paramterization of $Z_{\mathrm{P}}$, combined with eqs.~\eqref{eq:DefZM}
and~\eqref{eq:RegIndepPartZM}, provides
a parameterization of $Z_{\mathrm{M}}$ itself.  It has the form
\begin{equation}
\begin{split}
\label{eq:ZMbeta}
&Z_{\mathrm{M}}(g_{0}) = \sum_{i=0}^{2} z_{i}^{\textrm{M}}(\beta-6.0)^{i} \, ,\\
&\beta = 6/g_{0}^{2}, \quad 6.0 \le \beta \le 6.5 \, ,
\end{split}
\end{equation}
with the coefficients presented in tab.~\ref{tab:ZPbetaNP}.
The uncertainty of $M/\overline{\mu}_q(L)$ is independent of $\beta$ and
hence can be accounted for after the extrapolation to the continuum
limit has been carried out.

From eq.~\eqref{eq:ZPmatchingbeta} and eq.~\eqref{eq:ZMbeta},
it is now possible to compute $Z_{\mathrm{P}}$ and
$Z_{\mathrm{M}}$ at any value of $\beta$ within the range
$6.0 \le \beta \le 6.5$. This is 
the range of $\beta$ where large volume calculations have been performed, namely
$\beta = 6.00, 6.10,6.20,6.45$. For these computations, a number of bare quark masses were
used, which we can use to determine the renormalized quark mass from our
knowledge of $\zp$ and $Z_{\rm M}$.
The relevant values of $Z_{\mathrm{P}}$ and $Z_{\mathrm{M}}$ at
the chosen values of $\beta$ are summarized in tab.~\ref{tab:ZMbetainterest}.

As we just did for the pseudoscalar density, we 
now determine the RGI Z-factors of the
operators $O_{12}^{a}$ and $O_{44}^{a}$ using the $\chi$SF formulation.
These factors then relate the bare and the RGI matrix elements of the
corresponding operator.

\begin{figure}[t]
\centering
\includegraphics[width=0.8\textwidth]{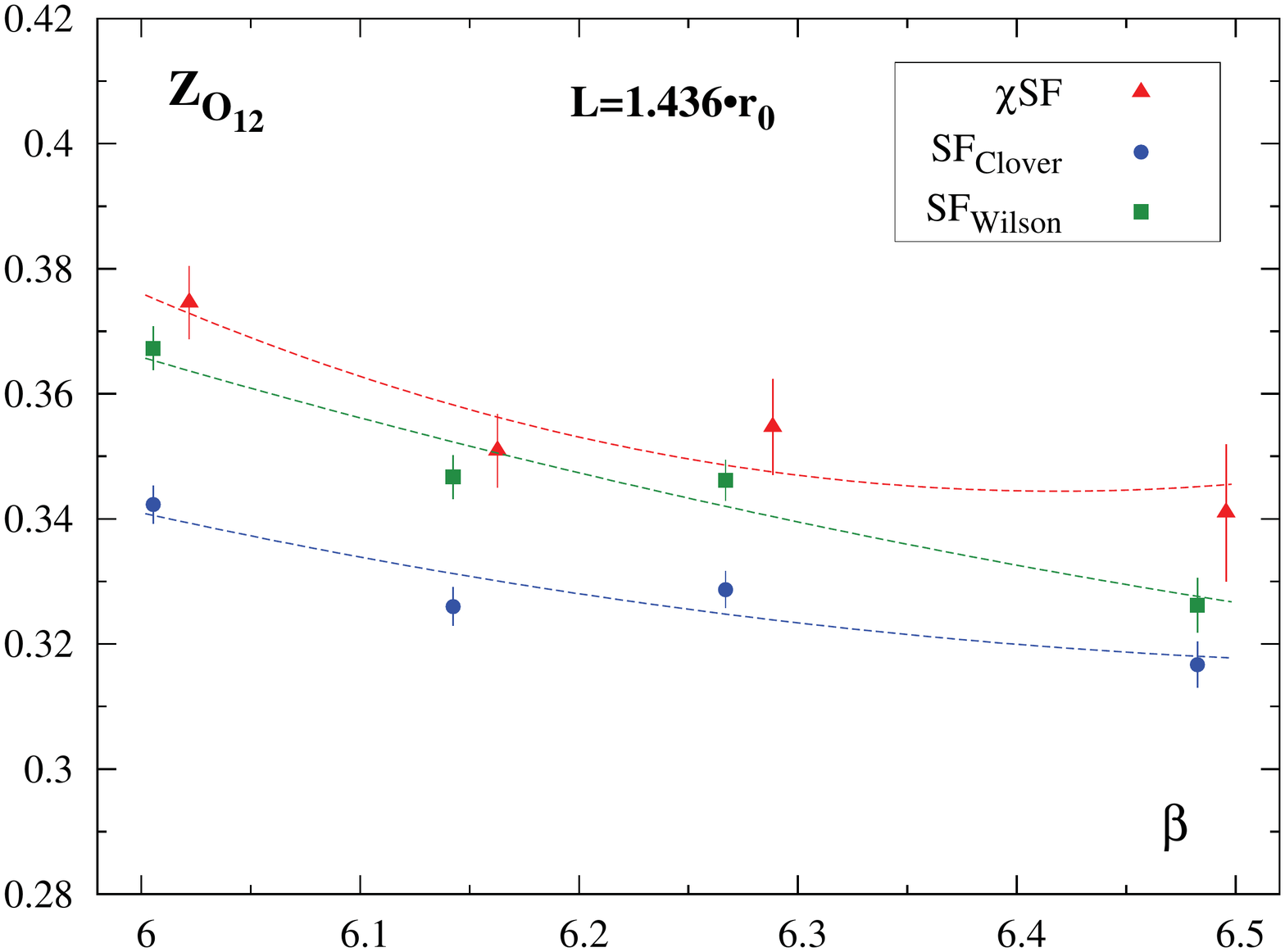}
\caption{Results for $Z_{\mathrm{O}_{12}}(g_{0},L/a)$ at the
  scale $L=1.436\,r_{0}$, $\boldsymbol{\theta}=(1,0,0)$ and for several values of $\beta$.
  Results are shown for the $\chi$SF with standard Wilson fermions
  (cf. tab.~\ref{tab:ZO12})
  and for the SF with standard and improved Wilson fermions, as taken
  from~\cite{Guagnelli:2004ga}.
  The fitting curves are also plotted
  (cf. eq.~\eqref{eq:ZOmatchingbeta} and tab.~\eqref{tab:ZObetaNP}).}
\label{fig:ZO12betaNPTheta100}
\end{figure}

\begin{figure}
\centering
\includegraphics[width=0.8\textwidth]{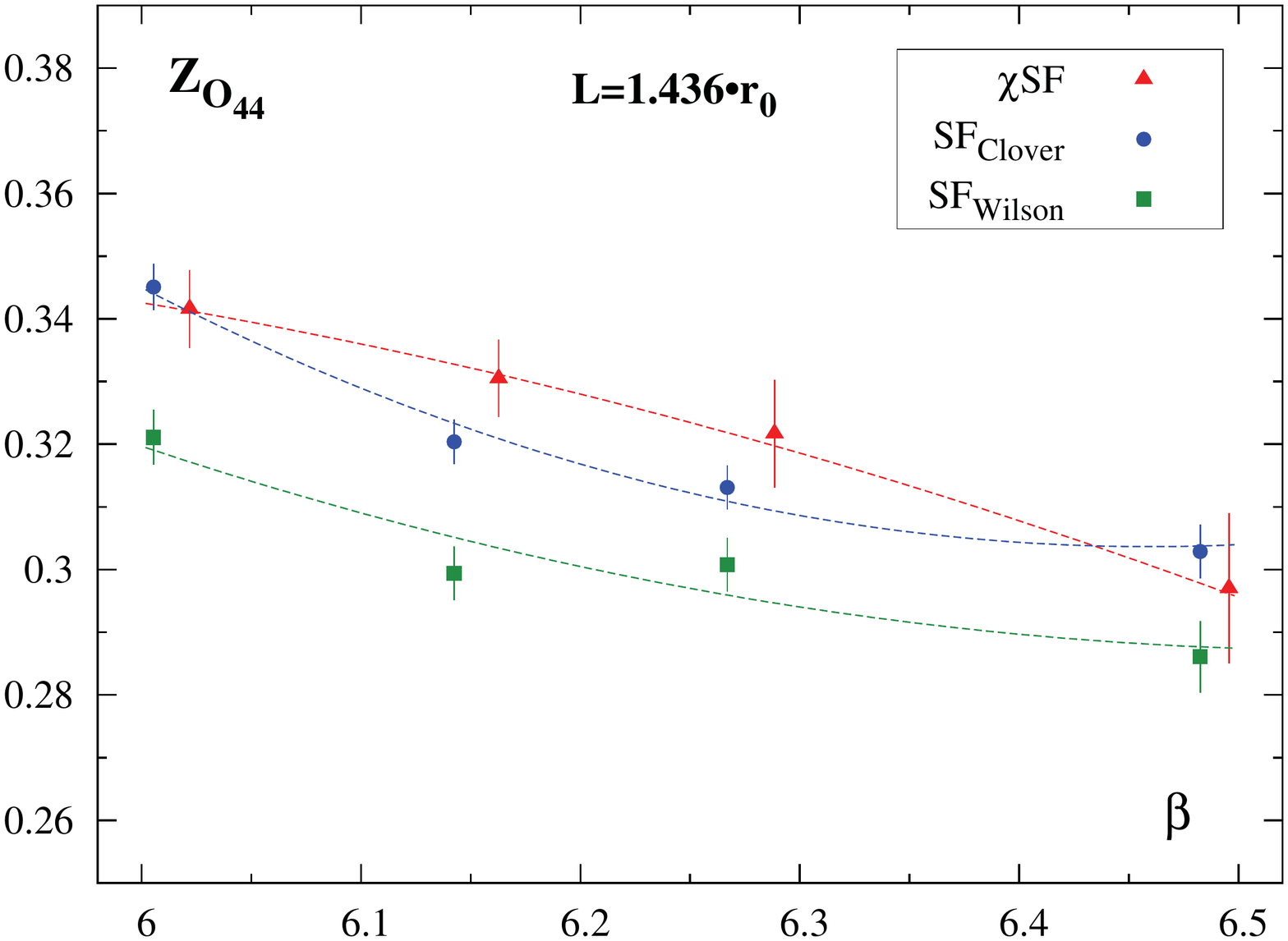}
\caption{ Results for $Z_{\mathrm{O}_{44}}(g_{0},L/a)$ at the
  scale $L=1.436\,r_{0}$, $\boldsymbol{\theta}=(1,0,0)$ and for several values of $\beta$.
  This is the same as fig.~\ref{fig:ZO12betaNPTheta100} but now for $O_{44}$
  instead of $O_{12}$.}
\label{fig:ZO44betaNPTheta100}
\end{figure}

We first study the dependence of the $Z$-factors on $\beta$
at the matching scale $L=1.436\, r_{0}$
and determine a curve describing this dependence.
Performing a fit of the values for the $Z$-factors
(found in tabs.~\ref{tab:ZO12} and~\ref{tab:ZO44})
of the form,
\be
\label{eq:ZOmatchingbeta}
\begin{split}
&Z_{\mathrm{O}}(g_{0},L/a)_{L=1.436\, r_{0}} =
\sum_{i=0}^{2} z_{i}^{\mathrm{REN}}(\beta-6.0)^{i} \, ,\\
&\beta = 6/g_{0}^{2}, \qquad 6.0 \le \beta \le 6.5 \, ,
\end{split}
\ee
we obtain the fitting coefficients $z_{i}^{\mathrm{REN}}$ given in
tab.~\ref{tab:ZObetaNP} for the $\chi$SF and the SF with standard and
improved Wilson fermions.
In eq.~\eqref{eq:ZOmatchingbeta}, `REN' stands for the
particular setup chosen: $\chi$SF, SF with standard
Wilson fermions or SF with improved Wilson fermions.
We have computed the Z-factors for $O_{12}^{a}$ and $O_{44}^{a}$ at $\boldsymbol{\theta}=(1,0,0)$
for all formulations and at $\boldsymbol{\theta}=(0.5,0.5,0.5)$ only for the
$\chi$SF.
These results are obtained from fits performed in this work for the
three formulations,
which for the SF are in agreement with the final results previously
presented in~\cite{Guagnelli:2004ga}.
We show the Z-factors together with the fitting curves,
at $\boldsymbol{\theta}=(1,0,0)$ and for the three formulations, in
figs.~\ref{fig:ZO12betaNPTheta100} and~\ref{fig:ZO44betaNPTheta100}
for $Z_{\mathrm{O}_{12}}$ and $Z_{\mathrm{O}_{44}}$, respectively.

As for the case of the pseudoscalar density, it is possible 
to determine the RGI renormalization constants for the twist-2 operators
from the knowledge of $Z^{\mathrm{REN}}(g_{0},L/a)$ 
at a given value of the renormalization scale, $1/L$.
For the twist-2 operators one defines, at the same value of the renormalization scale,
the ultraviolet (UV) invariant SSF,
$\sigma_{\mathrm{INV},\mathrm{O}}^{\mathrm{UV},\mathrm{REN}}(L)$.
This SSF is discussed thoroughly in
ref.~\cite{Guagnelli:1999wp}, which we assume some familiarity with.

The UV invariant SSF of a certain operator is independent of the
particular regularization but it depends on the renormalization scheme
and the matching scale.
In particular it is defined by
\begin{equation}
\label{eq:UVSSFDef}
O^{\mathrm{RGI}} = \sigma_{\mathrm{INV},\mathrm{O}}^{\mathrm{UV},\mathrm{REN}}(L)~O_{\mathrm{R}}(L) \, ,
\end{equation}
with $O_{\mathrm{R}}(L)$ the renormalized operator at
scale $1/L$, defined in eq.~\eqref{eq:RenOLat},
and the corresponding RGI operator $O^{\mathrm{RGI}}$.
We note that the UV invariant SSF is analogous 
to the factor $\left(M/\overline{\mu}_q(L)\right)$
that we have used for the RGI quark mass.

The RGI renormalization factor is scale and scheme independent
but it depends on the particular regularization.
It relates any bare matrix element of the bare operator,
$O_{\mathrm{B}}(g_{0})$,
with the corresponding RGI matrix element
and it is defined as follows,
\begin{equation}
\label{eq:DefZRGI}
Z_{\mathrm{O}}^{\mathrm{RGI}}(g_{0}) =
\frac{Z^{\mathrm{REN}}(g_{0},L/a)}
{\sigma_{\mathrm{INV},\mathrm{O}}^{\mathrm{UV},\mathrm{REN}}(L)} \, .
\end{equation}

In~\cite{Guagnelli:2003hw}, the value of the UV invariant SSF was given
for the operators $O_{12}^{a}$ and $O_{44}^{a}$ at the scale $L=1.436\,r_{0}$ and
for $\boldsymbol{\theta}=(1,0,0)$. The values given there are
\begin{equation}
\label{eq:RGISSFO}
\sigma_{\mathrm{INV},\mathrm{O}_{12}}^{\mathrm{UV},\mathrm{SF}} = 0.242\,(8) \, ,
\qquad
\sigma_{\mathrm{INV},\mathrm{O}_{44}}^{\mathrm{UV},\mathrm{SF}} = 0.221\,(9) \, .
\end{equation}

Substituting these values into eq.~\eqref{eq:DefZRGI} 
and using the Z-factors, at the matching scale
$L=1.436\,r_{0}$, in
tabs.~\ref{tab:ZO12} and~\ref{tab:ZO44}, 
the RGI Z-factors of the operators
$O_{12}^{a}$ and $O_{44}^{a}$ are determined and given in
tab.~\ref{tab:ZORGINPTheta100}. Results are provided only for
$\boldsymbol{\theta}=(1,0,0)$ because the UV invariant SSFs are only known 
for that case.
In the determination of the RGI Z-factors,
the uncertainty in the UV invariant SSF is not taken into
account. This is a quantity in the continuum, and therefore its uncertainty
is only considered at the end of all calculations, after the continuum
limit has been performed. Its error is then added in quadrature to
the final uncertainty in the continuum limit.

We smoothly parameterize $Z_{\mathrm{O}}^{\mathrm{RGI}}(g_{0})$ as a function of
$\beta$ by fitting the values in
tab.~\ref{tab:ZORGINPTheta100} to the following functional form
\begin{equation}
\label{eq:ZORGImatchingbeta}
\begin{split}
&Z_{\mathrm{O}}^{\mathrm{RGI}}(g_{0}) =
\sum_{i=0}^{2} z_{i}^{\mathrm{RGI}}(\beta-6.0)^{i} \, ,\\
&\beta = 6/g_{0}^{2}, \qquad 6.0 \le \beta \le 6.5 \,.
\end{split}
\end{equation}
The fitted coefficients can be found in tab.~\ref{tab:ZORGIbetaNP}.

Using the parameterizations in eqs.~\eqref{eq:ZOmatchingbeta} and~\eqref{eq:ZORGImatchingbeta}, 
both the Z-factors and the RGI Z-factors can
be determined at any value of $\beta$ within the range $6.0 \le \beta \le 6.5$.
In tab.~\ref{tab:ZObetainterest}, we provide results
for the particular values of
$\beta$ for which
bare matrix elements have
been evaluated in large volume
calculations~\cite{Capitani:2005jp}.

The results presented in
tab.~\ref{tab:ZObetainterest}
correspond to
$Z_{\mathrm{O}_{12}}$, $Z_{\mathrm{O}_{12}}^{\mathrm{RGI}}$, $Z_{\mathrm{O}_{44}}$
and $Z_{\mathrm{O}_{44}}^{\mathrm{RGI}}$
at $\boldsymbol{\theta}=(1,0,0)$.
We also give there the corresponding
results for the SF formulation with improved and standard Wilson
fermions.
Note that these values should not be compared across the three formulations.
They depend on the regularization.
Only a comparison of renormalized matrix
elements in the continuum limit would make sense.

Nevertheless, this calculation of the renormalization constants demonstrates that the
$\chi$SF can indeed be used in pratice to non-perturbatively renormalize
challenging operators, such as the twist-two operators considered in
this work.

\section{Strange quark mass}
\label{sec:ms}

In this section we compute the RGI strange quark mass, $M_{\mathrm{s}}$,
and the running strange quark mass in the $\overline{\mathrm{MS}}$-scheme at $2$~GeV
in quenched QCD. We use the $\chi$SF renormalization scheme with the
setup discussed in the previous sections
together with the bare quark masses from large volume calculations with twisted mass
fermions at maximal twist.

The purpose of this computation is to perform another check
of the $\chi$SF formulation.
In practice, we compute the quantity
$r_{0}\,(M_{\mathrm{s}}+\hat{M})$,
where $\hat{M}=(M_{\mathrm{u}}+M_{\mathrm{d}})/2$ is the average light quark mass and $r_0$ is the Sommer
parameter~\cite{Sommer:1993ce}.
We then take the continuum limit.
The resulting continuum limit value, obtained from the $\chi$SF,
is compared to that obtained from the standard SF with improved 
Wilson fermions~\cite{Garden:1999fg}.
We find that the two results agree, which is another check of universality in the
continuum limit.

Moreover, $r_{0}\,(M_{\mathrm{s}}+\hat{M})$ is expected to scale towards the continuum
limit with leading $O(a^{2})$ discretization errors,
up to possible boundary effects.
In fact, we will show that the scaling behavior is
consistent with leading $O(a^{2})$ discretization effects.
This represents another test of bulk automatic $O(a)$ improvement and,
moreover, it provides another indirect indication that the boundary effects coming from
$d_{s}$ (see ref.~\cite{Lopez:2012as}) are negligible, even at the large values of $g_{0}$ considered
in this section.

To determine the strange quark mass, we follow the strategy of ref.~\cite{Garden:1999fg}
and determine a reference bare quark mass $\mu_{\mathrm{ref}}$ defined by
$2\mu_{\mathrm{ref}} = \mu_{\mathrm{s}} + \hat{\mu}$,
where $\hat{\mu}=(\mu_{\mathrm{u}}+\mu_{\mathrm{d}})/2$.
The reference quark mass is then chosen such that the physical value of the 
kaon meson mass is reproduced.

To start, we determine the bare reference quark mass in lattice
units, $a\mu_{\mathrm{ref}}(g_{0})$, using the results of the large volume calculations
in~\cite{Jansen:2005kk}.  The pseudoscalar mass, $m_{\mathrm{PS}}$, was computed there in
lattice units using twisted mass Wilson fermions at maximal twist.  We
focus on the values of $\beta$ in~\cite{Jansen:2005kk} that overlap with the range
covered in this work $6.0 \le \beta \le 6.5$.

The pseudoscalar mass range covered by these data set is
$270~\mathrm{MeV} < m_{\mathrm{PS}} < 1180~\mathrm{MeV}$.
Within this range, we may interpolate in the bare quark mass $a\mu_q$
at the experimental value of the kaon mass, $m_{\mathrm{K}}$.
At each value of $\beta$, we perform a quadratic interpolation in 
the bare quark mass. We have cross checked that a linear interpolation
with the 3 data points closest to the interpolation point gives consistent results.
The lattice spacing in physical
units is obtained using the $\beta$ dependence of $r_0/a$ 
from ref.~\cite{Guagnelli:1998ud}.
The final results for $\mu_{\textrm{ref}}\, r_{0}$,
together with the corresponding values of $r_{0}/a$ and
$a\mu_{\mathrm{ref}}$,
are summarized in tab.~\ref{tab:murefr0}.

\begin{table}
\centering
\begin{tabular}[c]{|c|l|l|l|}\hline
\multicolumn{1}{|c|}{$\beta$} &
\multicolumn{1}{|c|}{$r_{0}/a$} &
\multicolumn{1}{|c|}{$a\mu_{\textrm{ref}}$} &
\multicolumn{1}{|c|}{$\mu_{\textrm{ref}}\,r_{0}$} \\
\hline
\multicolumn{4}{|c|}{$\kcr^{\mathrm{pion}}$ definition}\\
\hline
6.00 & 5.368\,(22) & 0.01450\,(59)& 0.0778\,(32)\\
6.10 & 6.324\,(28) & 0.01216\,(40)& 0.0769\,(26)\\
6.20 & 7.360\,(35) & 0.01030\,(34)& 0.0758\,(25)\\
6.45 & 10.458\,(58) & & \\
\hline
\multicolumn{4}{|c|}{$\kcr^{\mathrm{PCAC}}$ definition}\\
\hline
6.00 & 5.368\,(22) & 0.01443\,(51)& 0.0775\,(28)\\
6.20 & 7.360\,(35) & 0.01029\,(27)& 0.0757\,(20)\\
\hline
\end{tabular}
\caption{$\mu_{\textrm{ref}}\,r_{0}$ at the value of the kaon mass and for
  particular values of $\beta$. Results are shown for both the pion and the
  PCAC definitions of the critical mass, as discussed in ref.~\cite{Jansen:2005kk}.}
\label{tab:murefr0}
\end{table}

\begin{figure}
\centering
\includegraphics[width=1.0\textwidth]{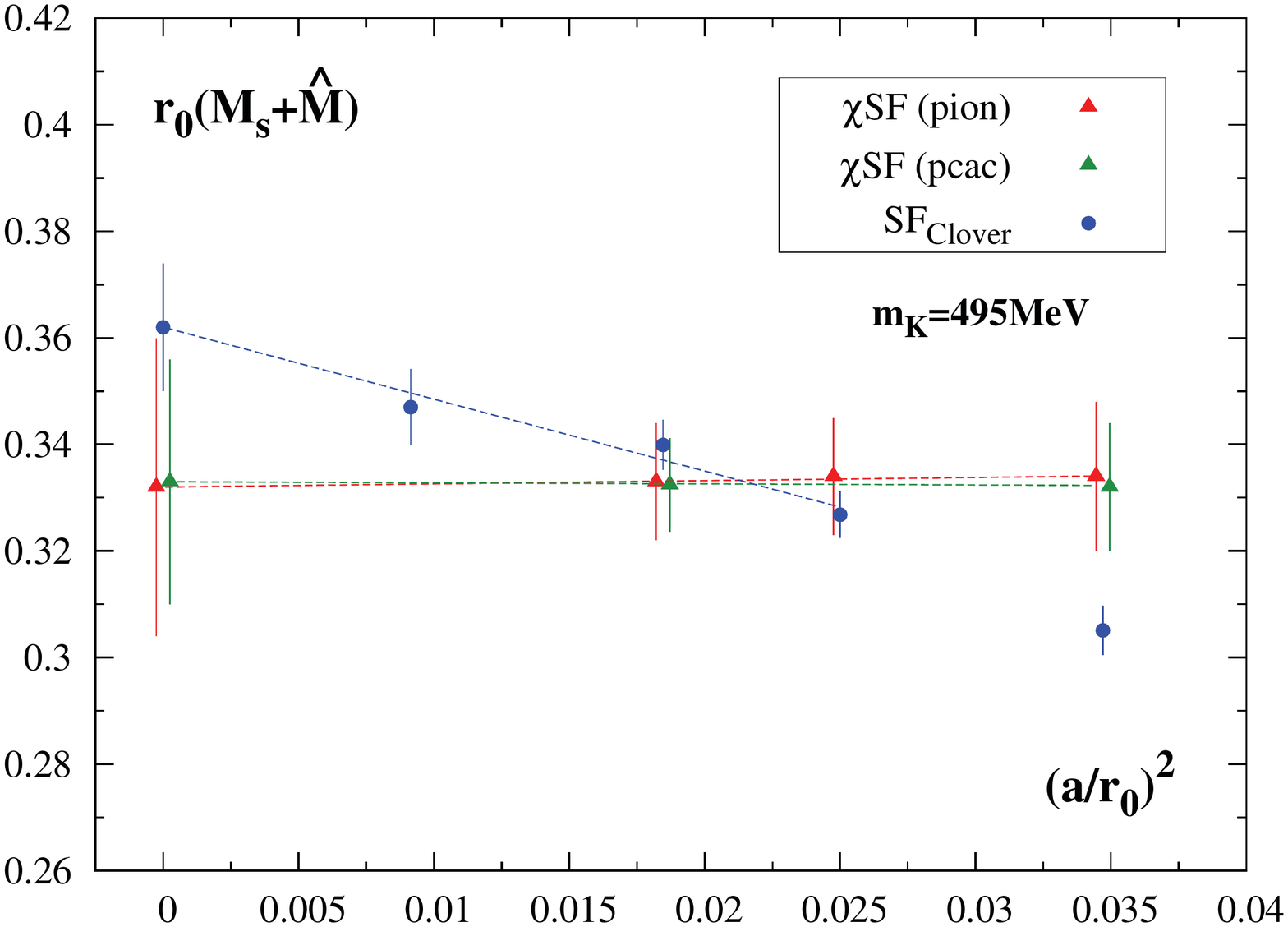}
\caption{Scaling of $r_{0}(M_{\mathrm{s}}+\hat{M})$ 
  at the physical value of the kaon mass, $m_{\mathrm{K}}=495~\mathrm{MeV}$.
  The extrapolations to the continuum limit are performed with
  linear fits in $(a/r_{0})^{2}$.
  The values in the continuum limit are also plotted.
  Results are shown for the $\chi$SF with standard Wilson fermions, for
  the two definitions of the critical mass, and also for the SF with
  improved Wilson fermions~\cite{Garden:1999fg}.
  The results for the $\chi$SF have been plotted slightly displaced to the right and
  left, respectively, for the PCAC and pion definitions of the critical mass.}
\label{fig:MK}
\end{figure}

Using the results for $Z_{\rm M}(g_0)$ obtained in sec.~\ref{sec:zfactors} and the 
reference quark mass just discusssed, we can determine the RGI strange quark mass.
The resulting values at finite lattice spacing for the RGI reference quark mass 
are summarized in tab.~\ref{tab:MKr0} and plotted in fig.~\ref{fig:MK}.

\begin{table}
\centering
\begin{tabular}[c]{|l|l|l|}\hline
\multicolumn{1}{|c|}{$\beta$} &
\multicolumn{1}{|c|}{$Z_{\mathrm{M}}$ } &
\multicolumn{1}{|c|}{$r_{0}\,(M_{\mathrm{s}}+\hat{M})$ } \\
\hline
\multicolumn{3}{|c|}{$\kcr^{\mathrm{pion}}$ definition}\\
\hline
6.00 & 2.1444 (55) & 0.334\,(14) \\
6.10 & 2.1733 (33) & 0.334\,(11) \\
6.20 & 2.1957 (42) & 0.333\,(11) \\
6.45 & 2.2236 (70) & \\
\hline
\multicolumn{3}{|c|}{$\kcr^{\mathrm{PCAC}}$ definition}\\
\hline
6.00 & 2.1444 (55) & 0.332\,(12) \\
6.20 & 2.1957 (42) & 0.3324\,(88)\\
\hline
\end{tabular}
\caption{$r_{0}\,(M_{\mathrm{s}}+\hat{M})=Z_{\mathrm{M}}(2\mu_{\textrm{ref}}\,r_{0})$ at
  several values of $\beta$. Results are shown for the pion and the
  PCAC definitions of the critical mass.}
\label{tab:MKr0}
\end{table}

We plot the results for $r_{0}\,(M_{\mathrm{s}}+\hat{M})$ 
obtained using the two methods for determining $\kcr$ 
in ref.~\cite{Jansen:2005kk}. 
The subtleties related to these 2 different choices are discussed
in ref.~\cite{Shindler:2007vp} and refs. therein. For this work, these two definitions
simply correspond to two slighlty different discretizations of the twisted mass action 
inducing slightly different O($a^2$) cutoff effects in the physical quantities.

In fig.~\ref{fig:MK} we also show a continuum extrapolation, linear in $(a/r_{0})^{2}$, 
and the resulting continuum limit values.
The results obtained from the SF with improved Wilson
fermions~\cite{Garden:1999fg} are also plotted for comparison. 
The final values in the continuum limit
are given in eqs.~\eqref{eq:MKpion} and~\eqref{eq:MKpcac}
for the pion and PCAC definitions of the critical mass, 
\begin{align}
\mathrm{pion:} \quad r_{0}\,(M_{\mathrm{s}}+\hat{M})^{\chi \mathrm{SF}} &= 0.332\,(28) \, , \label{eq:MKpion}\\
\mathrm{PCAC:} \quad r_{0}\,(M_{\mathrm{s}}+\hat{M})^{\chi \mathrm{SF}} &= 0.333\,(23) \, . \label{eq:MKpcac}
\end{align}
The errors include the uncertainty of $M/\overline{\mu}_q(L)$
added in quadrature.
These values are consistent with that obtained in~\cite{Garden:1999fg}
\begin{equation}
\label{eq:MKalpha}
r_{0}\,(M_{\mathrm{s}}+\hat{M})^{\textrm{SF}} = 0.362\,(12) \, .
\end{equation}
As can be seen from these results and those in
fig.~\ref{fig:MK}, the SF values have relative errors that are about
two times smaller than those obtained in our calculation using
the $\chi$SF. 
This difference is due to the size of the statistical uncertainties in
the bare pseudoscalar masses obtained from the corresponding
large volume calculations and not due to an inherit difference in
the accuracies that can be obtained with $\chi$SF or SF computations.
Our $\chi$SF values rely on the results of~\cite{Jansen:2005kk} 
whereas the SF values use those in~\cite{Garden:1999fg}, 
which are about twice as accurate as the values in~\cite{Jansen:2005kk}. 
This accounts for the larger uncertainty of the $\chi$SF results
for the strange quark mass.

We can conclude that the values of the RGI reference quark mass,
and therefore the RGI strange quark mass itself,
determined using the SF and the $\chi$SF agree in the continuum limit.
This is another test of the universality,
this time at a rather large value of the physical volume.
In particular, these results demonstrate that the $\chi$SF, like the SF, could also be a
valuable tool for the computation of the renoramalized quark
masses. 

Even if not necessary for testing universality,
we compute, for completeness, the values of both the RGI
strange quark mass in physical units and the
running strange quark mass in the $\overline{\mathrm{MS}}$-scheme.
As discussed in~\cite{Garden:1999fg},
chiral perturbation theory allows for a precise determination of
ratios of the masses of the three lightest 
quarks~\cite{Gasser:1982ap,Leutwyler:1994fi,Leutwyler:1996qg}.
Such determinations lead to
\begin{equation}
\label{eq:RatiosQuarkMassesChiPT}
M_{\mathrm{u}}/M_{\mathrm{d}} = 0.553 \pm 0.043 \, ,
\qquad
M_{\mathrm{s}}/\hat{M} = 24.4 \pm 1.5 \,.
\end{equation}

Assuming these relations together with eqs.~\eqref{eq:MKpion}
and~\eqref{eq:MKpcac}, we can determine the value of the RGI
strange quark mass. The final results in units of $r_{0}$ are
\begin{align}
\mathrm{pion:} \quad r_{0}\,M_{\mathrm{s}}^{\chi \mathrm{SF}}  &= 0.319\,(27) \, , \label{eq:Mspion} \\
\mathrm{PCAC:} \quad r_{0}\,M_{\mathrm{s}}^{\chi\mathrm{SF}} &= 0.320\,(22) \, , \label{eq:Mspcac}
\end{align}
for the pion and PCAC definitions of $\kcr$, respectively.
Repeating this analysis for the results in~\cite{Garden:1999fg} we obtain,
\begin{equation}
\label{eq:Msalpha}
r_{0}\,M_{\mathrm{s}}^{\mathrm{SF}} = 0.348\,(12) \, .
\end{equation}

The value of the RGI strange quark mass can be now given in physical
units,
\begin{align}
\mathrm{pion:} \quad  M_{\mathrm{s}}^{\chi \mathrm{SF}} &= 126\,(11)\,
\mathrm{MeV}\, , \label{eq:MspionPhysUnits} \\
\mathrm{PCAC:} \quad M_{\mathrm{s}}^{\chi\mathrm{SF}}  &= 126\,(9)\, \mathrm{MeV} \, , \label{eq:MspcacPhysUnits}
\end{align}
and for the SF,
\begin{equation}
\label{eq:MsalphaPhysUnits}
M_{\mathrm{s}}^{\mathrm{SF}} = 137\,(5)\, \mathrm{MeV} \, .
\end{equation}

Using the conversion factor between the RGI mass and the
running mass in the $\overline{\mathrm{MS}}$-scheme,
the running strange quark mass in the $\overline{\mathrm{MS}}$-scheme
can be determined.
At a value of the energy scale of $2$ GeV and up to $4$ loops,
the flavor-independent conversion factor is~\cite{Garden:1999fg}
\begin{equation}
\label{eq:ConversionMRGImmsbar}
\overline{m}^{\overline{\mathrm{MS}}}(2\, \mathrm{GeV})/M = 0.72076
\, .
\end{equation}
As a result,
the strange quark mass at 2 GeV with 4-loop running in the
$\overline{\mathrm{MS}}$-scheme is 
\begin{align}
\chi\textrm{SF:} \quad \overline{\mu}_{\mathrm{s}}^{\overline{\mathrm{MS}}}(2\, \mathrm{GeV})  &= 91\,(6)\, \mathrm{MeV} \, , \label{eq:msMSbarPcac}\\
\textrm{SF:} \quad \overline{m}_{\mathrm{s}}^{\overline{\mathrm{MS}}}(2\, \mathrm{GeV}) &= 99\,(4)\, \mathrm{MeV} \, , \label{eq:msMSbarAlpha}
\end{align}
where we quote as our best result the strange quark mass obtained from the PCAC definition of $\kcr$.

\section{Conclusions}
\label{sec:conclu}

Presently to renormalize Wilson twisted mass fermions preserving the property of 
automatic O($a$) improvement, infinite volume renormalization schemes such as the 
RI-MOM~\cite{Martinelli:1994ty,Constantinou:2010gr} or the 
X-space schemes~\cite{Gimenez:2004me,Cichy:2012is} are used.
However, only finite volume schemes such as the SF~\cite{Luscher:1992an,Jansen:1995ck} and the
$\chi$SF~\cite{Sint:2010eh} solve the problem of covering a large range of scales. 
Furthermore, the $\chi$SF scheme, when used to renormalize bare matrix elements computed
with maximally twisted mass fermions, is also compatible with automatic O($a$) improvement.

In this work we have made a detailed study of several applications of 
the $\chi$SF scheme, with quenched Wilson fermions. 
Using our results for the non-perturbative tuning of the $\chi$SF~\cite{Lopez:2012as},
we have performed a number of scaling studies of the $\chi$SF. 
We have analyzed step-scaling functions of the pseudoscalar 
density and of two discretizations of a twist-2 operator at a perturbative and an intermediate value
of the renormalized coupling. All our results agree in the continuum
limit with those obtained from the standard SF scheme.
Additionally, our results are consistent with scaling violations of only O($a^2$), thus demonstrating
the bulk O($a$) improvement of the $\chi$SF scheme.

We remark that for the twist-2 operators this is an important result because 
to improve such operators in the standard SF scheme would require a non-perturbative 
computation of additional improvement coefficients.
The automatic O($a$) improvement found here with the example
of the twist-2 operators can be taken over to other observables.
It demonstrates that automatic O($a$) improvement is at work and that with the $\chi$SF scheme,
in combination with maximally twisted mass fermions, the somewhat demanding computation of operator
specific improvement coefficients can be avoided.

Additionally, we have computed the continuum limit of the renormalized strange quark mass within
the $\chi$SF scheme. In this case, as well, the result in the continuum limit is consistent with 
previous results obtained with non-perturbatively improved Wilson fermions and the standard SF scheme,
thus demonstrating that the $\chi$SF scheme works as well as the standard SF scheme even for
the more commonly computed quantities, such as the running of the
quark masses.

Therefore, we believe
that the $\chi$SF scheme is both a practical and theoretically well defined framework
that can be used as a non-perturbative renormalization scheme for large volume calculations 
of bare operators.

\section*{Acknowledgments}
\vspace{-1.0cm}

We thank S. Sint and B. Leder for many useful discussions. 
We also acknowledge the support of the computer center in DESY-Zeuthen and the NW-grid in Lancaster.
This work has been supported in part by the DFG Sonderforschungsbereich/Transregio SFB/TR9-03.
This manuscript has been coauthored by Jefferson Science Associates, LLC under Contract 
No. DE-AC05-06OR23177 with the U.S. Department of Energy.

\newpage

\begin{appendix}

\section{Tables of numerical results for the step-scaling functions}
\label{app:SSF}
\scriptsize

\begin{longtable}[c]{|r|l||l|l||l|l|}
\hline
\multicolumn{2}{|c||}{} &
\multicolumn{2}{|c||}{$\boldsymbol{\theta}=(0.5,0.5,0.5)$} &
\multicolumn{2}{|c|}{$\boldsymbol{\theta}=(1,0,0)$}\\
\hline
\multicolumn{1}{|c|}{$L/a$} &
\multicolumn{1}{|c||}{$\beta$} &
\multicolumn{1}{|c|}{$Z_{\textrm{P}}(g_{0},L/a)$ } &
\multicolumn{1}{|c||}{$Z_{\textrm{P}}(g_{0},2L/a)$ } &
\multicolumn{1}{|c|}{$Z_{\textrm{P}}(g_{0},L/a)$ } &
\multicolumn{1}{|c|}{$Z_{\textrm{P}}(g_{0},2L/a)$ } \\
\hline
\multicolumn{6}{|c|}{Hadronic scale: $L=1.436\, r_{0}$} \\
\hline
  8 & 6.0219 & 0.5385\,(12) &  & 0.5432\,(12) & \\
10 & 6.1628 & 0.5264\,(12) &  & 0.5310\,(12) & \\
12 & 6.2885 & 0.5272\,(16) &  & 0.5321\,(17) & \\
16 & 6.4956 & 0.5187\,(22) &  & 0.5245\,(21) & \\
\hline
\multicolumn{6}{|c|}{Intermediate scale: $\overline{g}^{2}=2.4484$} \\
\hline
  8 & 7.0197 & 0.68509\,(95) & 0.6199\,(14) & 0.68850\,(93) & 0.6241\,(13) \\
12 & 7.3551 & 0.6735 \,(13)  & 0.6082\,(19) & 0.6788\,(12)   & 0.6142\,(21) \\
16 & 7.6101 & 0.6672 \,(16)  & 0.5991\,(22) & 0.6737\,(16)   & 0.6015\,(22) \\
\hline
\multicolumn{6}{|c|}{Perturbative scale: $\overline{g}^{2}=0.9944$} \\
\hline
 8  & 10.3000 & 0.82689\,(56) & 0.80129\,(84) & 0.83007\,(58) & 0.80358\,(84) \\
12 & 10.6086 & 0.81651\,(88) & 0.78549\,(84) & 0.81924\,(82) & 0.79008\,(80) \\
16 & 10.8910 & 0.8110 \,(10)  & 0.7820\,(14)   & 0.8136\,(11)   & 0.7802\,(15)   \\
\hline
\caption{Renormalization factors of the pseudoscalar density,
  $Z_{\textrm{P}}$, at $\boldsymbol{\theta}=(0.5,0.5,0.5)$ and
  $\boldsymbol{\theta}=(1,0,0)$.
  Results are shown for the $\chi$SF with standard Wilson fermions
  at three values of the renormalization scale and for several values
  of the lattice spacing.}
\label{tab:ZP}
\end{longtable}

\begin{minipage}[b]{0.5\linewidth}
\begin{longtable}[c]{|r|c|c|c|}
\hline
\multicolumn{4}{|c|}{$\Sigma_{\textrm{P}}(2,u,a/L)$} \\\hline
\multicolumn{1}{|c|}{$L/a$} &
\multicolumn{1}{|c|}{$\chi$SF} &
\multicolumn{1}{|c|}{SF (Clover)} &
\multicolumn{1}{|c|}{SF (Wilson)} \\ \hline
\multicolumn{4}{|c|}{Intermediate scale: $\overline{g}^{2}=2.4484$} \\\hline
  8 & 0.9048\,(23) & 0.8945\,(23) & 0.8993\,(20) \\
12 & 0.9030\,(33) & 0.8908\,(23) & 0.8924\,(30) \\
16 & 0.8980\,(39) & 0.8998\,(25) & 0.9036\,(32) \\
\hline
\multicolumn{4}{|c|}{Perturbative scale: $\overline{g}^{2}=0.9944$} \\\hline
 8 & 0.9690\,(12) & 0.9633\,(14) & 0.9641\,(12) \\
12 & 0.9620\,(15) & 0.9599\,(19) & 0.9632\,(17) \\
16 & 0.9643\,(22) & 0.9622\,(20) & 0.9652\,(22) \\
\hline
\caption{SSF of the pseudoscalar density at finite lattice spacing,
  $\Sigma_{\textrm{P}}(2,u,a/L)$, for $\boldsymbol{\theta}=(0.5,0.5,0.5)$.
  Results are shown for the $\chi$SF with standard Wilson fermions
  and also for the SF with improved and standard
  Wilson fermions~\cite{Guagnelli:2004za} at two values of the
  renormalization scale and for several values of the lattice spacing.}
\label{tab:LSSFTheta0.5}
\end{longtable}
\end{minipage}
\hspace{0.5cm}
\begin{minipage}[b]{0.5\linewidth}
\begin{longtable}[c]{|r|c|}
\hline
\multicolumn{2}{|c|}{$\Sigma_{\textrm{P}}(2,u,a/L)$} \\\hline
\multicolumn{1}{|c|}{$L/a$} &
\multicolumn{1}{|c|}{$\chi$SF} \\ \hline
\multicolumn{2}{|c|}{Intermediate scale: $\overline{g}^{2}=2.4484$} \\\hline
  8 & 0.9065\,(23) \\
12 & 0.9048\,(35) \\
16 & 0.8929\,(39) \\
\hline
\multicolumn{2}{|c|}{Perturbative scale: $\overline{g}^{2}=0.9944$} \\\hline
  8 & 0.9681\,(12) \\
12 & 0.9644\,(14) \\
16 & 0.9590\,(23) \\
\hline
\caption{SSF of the pseudoscalar density at finite lattice spacing,
  $\Sigma_{\textrm{P}}(2,u,a/L)$, for $\boldsymbol{\theta}=(1,0,0)$.
  Results are shown for the $\chi$SF with standard Wilson fermions
  at two values of the renormalization scale and for several values of
  the lattice spacing.}
\label{tab:LSSFTheta100}
\end{longtable}
\end{minipage}

\begin{minipage}[b]{0.5\linewidth}
\begin{longtable}[c]{|c|c|c|c|}
\hline
\multicolumn{1}{|c|}{} &
\multicolumn{1}{|c|}{$\chi$SF} &
\multicolumn{1}{|c|}{SF (Clover)}  &
\multicolumn{1}{|c|}{SF (Wilson)} \\\hline
\multicolumn{4}{|c|}{Intermediate scale: $\overline{g}^{2}=2.4484$} \\ \hline
$\sigma_{\textrm{P}}(2,u)$ & 0.8981\,(41)  & 0.8968\,(28) & 0.8993\,(58)\\
slope                                & 0.44\,(34)  &   -0.22\,(27) & -0.007\,(56)\\
$\chi^{2}/\mathrm{dof}$                  & 0.5349  &          6.4083 &          6.8156\\
\hline
\multicolumn{4}{|c|}{Perturbative scale: $\overline{g}^{2}=0.9944$} \\\hline
$\sigma_{\textrm{P}}(2,u)$ & 0.9595\,(21) & 0.9602\,(22) & 0.9644\,(37)\\
slope                                & 0.59\,(18) &     0.19\,(19) & -0.004\,(35)\\
$\chi^{2}/\mathrm{dof}$                  & 2.4748 &          1.1258 &          0.5129\\
\hline
\caption{Continuum limit of the SSF of the pseudoscalar density.
  Results are shown for the $\chi$SF with standard Wilson fermions
  and also for the SF with improved and standard Wilson fermions
  at two values of the renormalization scale and for $\boldsymbol{\theta}=(0.5,0.5,0.5)$.
  These results correspond to linear fits of the values in
  tab.~\ref{tab:LSSFTheta0.5}. The fit is linear in $a/L$ for the
  SF(Wilson) formulation while it is linear in $(a/L)^{2}$ for the
  $\chi$SF and SF(Clover) formulations.}
\label{tab:SSFcont.Theta0.5}
\end{longtable}
\end{minipage}
\hspace{1.0cm}
\begin{minipage}[b]{0.5\linewidth}
\begin{longtable}[c]{|c|c|}
\hline
\multicolumn{1}{|c|}{} &
\multicolumn{1}{|c|}{$\chi$SF} \\ \hline
\multicolumn{2}{|c|}{Intermediate scale: $\overline{g}^{2}=2.4484$} \\ \hline
$\sigma_{\textrm{P}}(2,u)$ & 0.8942\,(42)\\
slope                                & 0.82\,(34)\\
$\chi^{2}/\mathrm{dof}$                  & 3.3471 \\
\hline
\multicolumn{2}{|c|}{Perturbative scale: $\overline{g}^{2}=0.9944$} \\\hline
$\sigma_{\textrm{P}}(2,u)$ & 0.9591\,(21)\\
slope                                & 0.59\,(17) \\
$\chi^{2}/\mathrm{dof}$                  & 1.8644 \\
\hline
\caption{Continuum limit of the SSF of the pseudoscalar density.
  Results are shown for the $\chi$SF with standard Wilson fermions
  at two values of the renormalization scale and for $\boldsymbol{\theta}=(1,0,0)$.
  These results correspond to linear fits in $(a/L)^{2}$ of the values
  in tab.~\ref{tab:LSSFTheta100}.}
\label{tab:SSFcont.Theta100}
\end{longtable}
\end{minipage}

\begin{longtable}[c]{|r|l||l|l||l|l|}
\hline
\multicolumn{2}{|c||}{} &
\multicolumn{2}{|c||}{$\boldsymbol{\theta}=(0.5,0.5,0.5)$} &
\multicolumn{2}{|c|}{$\boldsymbol{\theta}=(1,0,0)$}\\
\hline
\multicolumn{1}{|c|}{$L/a$} &
\multicolumn{1}{|c||}{$\beta$} &
\multicolumn{1}{|c|}{$Z_{\textrm{O}_{12}}(g_{0},L/a)$ } &
\multicolumn{1}{|c||}{$Z_{\textrm{O}_{12}}(g_{0},2L/a)$ } &
\multicolumn{1}{|c|}{$Z_{\textrm{O}_{12}}(g_{0},L/a)$ } &
\multicolumn{1}{|c|}{$Z_{\textrm{O}_{12}}(g_{0},2L/a)$ } \\
\hline
\multicolumn{6}{|c|}{Hadronic scale: $L=1.436\, r_{0}$} \\
\hline
  8 & 6.0219 & 0.395\,(12) & & 0.3746\,(59) & \\
10 & 6.1628 & 0.374\,(13) & & 0.3509\,(59) & \\
12 & 6.2885 & 0.348\,(15) & & 0.3547\,(77) & \\
16 & 6.4956 & 0.353\,(21) & & 0.341\,(11)   &  \\
\hline
\multicolumn{6}{|c|}{Intermediate scale: $\overline{g}^{2}=2.4484$} \\
\hline
  8 & 7.0197 & 0.6077\,(80) & 0.482\,(13) & 0.5675\,(41) & 0.4498\,(64)\\
12 & 7.3551 & 0.613\,(11)   & 0.495\,(16) & 0.5634\,(58) & 0.4401\,(84)\\
16 & 7.6101 & 0.611\,(14)   & 0.460\,(16) & 0.5587\,(70) & 0.4383\,(84)\\
\hline
\multicolumn{6}{|c|}{Perturbative scale: $\overline{g}^{2}=0.9944$} \\
\hline
 8  & 10.3000 & 0.7989\,(44) & 0.7570\,(68) & 0.7717\,(25) & 0.7287\,(36)\\ 
12 & 10.6086 & 0.7800\,(66) & 0.7597\,(62) & 0.7530\,(35) & 0.7295\,(34)\\
16 & 10.8910 & 0.7762\,(83) & 0.730\,(11)   & 0.7511\,(45) & 0.7161\,(68)\\
\hline
\caption{Renormalization factors $Z_{\textrm{O}_{12}}$ for
  $\boldsymbol{\theta}=(0.5,0.5,0.5)$ and $\boldsymbol{\theta}=(1,0,0)$.
  Results are shown for the $\chi$SF with standard Wilson fermions
  at three values of the renormalization scale and for several values
  of the lattice spacing.}
\label{tab:ZO12}
\end{longtable}
\newpage
\begin{longtable}[c]{|r|l||l|l||l|l|}
\hline
\multicolumn{2}{|c||}{} &
\multicolumn{2}{|c||}{$\boldsymbol{\theta}=(0.5,0.5,0.5)$} &
\multicolumn{2}{|c|}{$\boldsymbol{\theta}=(1,0,0)$}\\
\hline
\multicolumn{1}{|c|}{$L/a$} &
\multicolumn{1}{|c||}{$\beta$} &
\multicolumn{1}{|c|}{$Z_{\textrm{O}_{44}}(g_{0},L/a)$ } &
\multicolumn{1}{|c||}{$Z_{\textrm{O}_{44}}(g_{0},2L/a)$ } &
\multicolumn{1}{|c|}{$Z_{\textrm{O}_{44}}(g_{0},L/a)$ } &
\multicolumn{1}{|c|}{$Z_{\textrm{O}_{44}}(g_{0},2L/a)$ } \\
\hline
\multicolumn{6}{|c|}{Hadronic scale: $L=1.436\, r_{0}$} \\
\hline
  8 & 6.0219 & 0.319\,(10) & & 0.3416\,(62) & \\
10 & 6.1628 & 0.307\,(10) & & 0.3305\,(62) & \\
12 & 6.2885 & 0.280\,(14) & & 0.3217\,(86) & \\
16 & 6.4956 & 0.261\,(19) & & 0.297\,(12)   & \\
\hline
\multicolumn{6}{|c|}{Intermediate scale: $\overline{g}^{2}=2.4484$} \\
\hline
  8 & 7.0197 & 0.5174\,(75) & 0.388\,(12) & 0.5382\,(44) & 0.4203\,(71)\\
12 & 7.3551 & 0.532\,(11)   & 0.404\,(16) & 0.5340\,(64) & 0.4189\,(93)\\
16 & 7.6101 & 0.521\,(14)   & 0.405\,(17) & 0.5236\,(82) & 0.4417\,(86)\\
\hline
\multicolumn{6}{|c|}{Perturbative scale: $\overline{g}^{2}=0.9944$} \\
\hline
 8  & 10.3000 & 0.7369\,(46) & 0.6781\,(71) & 0.7529\,(26) & 0.7044\,(39)\\
12 & 10.6086 & 0.7145\,(64) & 0.6882\,(68) & 0.7334\,(37) & 0.7068\,(39)\\
16 & 10.8910 & 0.7114\,(88) & 0.686\,(11)   & 0.7301\,(51) & 0.7135\,(65)\\
\hline
\caption{Renormalization factors $Z_{\textrm{O}_{44}}$ for
  $\boldsymbol{\theta}=(0.5,0.5,0.5)$ and $\boldsymbol{\theta}=(1,0,0)$.
  Results are shown for the $\chi$SF with standard Wilson fermions
  at three values of the renormalization scale and for several values
  of the lattice spacing.}
\label{tab:ZO44}
\end{longtable}
\begin{longtable}[c]{|r||r @{.} l||r @{.} l|}
\hline
\multicolumn{5}{|c|}{$\chi$SF} \\ \hline
\multicolumn{1}{|c||}{$L/a$} &
\multicolumn{2}{|c||}{$\Sigma_{\textrm{O}_{12}}(2,u,a/L)$} &
\multicolumn{2}{|c|}{$\Sigma_{\textrm{O}_{44}}(2,u,a/L)$} \\\hline
\multicolumn{5}{|c|}{Intermediate scale: $\overline{g}^{2}=2.4484$} \\\hline
  8 & 0&793\,(24) & 0&751\,(26) \\
12 & 0&809\,(31) & 0&761\,(34) \\
16 & 0&752\,(31) & 0&777\,(39) \\
\hline
\multicolumn{5}{|c|}{Perturbative scale: $\overline{g}^{2}=0.9944$} \\\hline
  8 & 0&948\,(10) & 0&920\,(11) \\
12 & 0&974\,(11) & 0&963\,(13) \\
16 & 0&940\,(18) & 0&964\,(20) \\
\hline
\caption{SSF of $O_{12}$ and $O_{44}$ at finite lattice spacing,
  $\Sigma_{\textrm{O}_{12}}$ and $\Sigma_{\textrm{O}_{44}}$.
  Results are shown for the $\chi$SF with standard Wilson fermions
  at two values of the renormalization scale, for several values of
  the lattice spacing and for $\boldsymbol{\theta}=(0.5,0.5,0.5)$.}
\label{tab:LSSFO12O44Theta0.5}
\end{longtable}
\newpage
\begin{longtable}[c]{|r||r @{.} l|r @{.} l|r @{.} l||r @{.} l|r @{.} l|r @{.} l|}
\hline
\multicolumn{1}{|c||}{} &
\multicolumn{6}{|c||}{$\Sigma_{\textrm{O}_{12}}(2,u,a/L)$} &
\multicolumn{6}{|c|}{$\Sigma_{\textrm{O}_{44}}(2,u,a/L)$} \\\hline
\multicolumn{1}{|c||}{$L/a$} &
\multicolumn{2}{|c|}{$\chi$SF} &
\multicolumn{2}{|c|}{SF (Clover)} &
\multicolumn{2}{|c||}{SF (Wilson)} &
\multicolumn{2}{|c|}{$\chi$SF} &
\multicolumn{2}{|c|}{SF (Clover)} &
\multicolumn{2}{|c|}{SF (Wilson)} \\ \hline
\multicolumn{13}{|c|}{Intermediate scale: $\overline{g}^{2}=2.4484$} \\\hline
  8 & 0&793\,(13) & 0&8223\,(77)   & 0&8811\,(85)   & 0&781\,(15) & 0&7885\,(91)   & 0&7935\,(119) \\
12 & 0&781\,(17) & 0&8053\,(77)   & 0&8589\,(136) & 0&784\,(20) & 0&7921\,(94)   & 0&7942\,(186) \\
16 & 0&785\,(18) & 0&8116\,(107) & 0&8519\,(85)   & 0&844\,(21) & 0&8036\,(127) & 0&7823\,(115) \\
\hline
\multicolumn{13}{|c|}{Perturbative scale: $\overline{g}^{2}=0.9944$} \\\hline
  8 & 0&9443\,(56) & \multicolumn{2}{|c|}{} & \multicolumn{2}{|c||}{} & 0&9355\,(62) & \multicolumn{2}{|c|}{} & \multicolumn{2}{|c|}{} \\
12 & 0&9688\,(64) & \multicolumn{2}{|c|}{} & \multicolumn{2}{|c||}{} & 0&9637\,(72) & \multicolumn{2}{|c|}{} & \multicolumn{2}{|c|}{} \\
16 & 0&953\,(11)   & \multicolumn{2}{|c|}{} & \multicolumn{2}{|c||}{} & 0&977\,(11)   & \multicolumn{2}{|c|}{} & \multicolumn{2}{|c|}{} \\
\hline
\caption{SSF of $O_{12}$ and $O_{44}$ at finite lattice spacing,
  $\Sigma_{\textrm{O}_{12}}$ and $\Sigma_{\textrm{O}_{44}}$.
  Results are shown for the $\chi$SF with standard Wilson fermions
  and also for the SF~\cite{Guagnelli:2003hw} with improved and
  standard Wilson fermions
  at two values of the renormalization scale, for several values of
  the lattice spacing and for $\boldsymbol{\theta}=(1,0,0)$.}
\label{tab:LSSFO12O44Theta100}
\end{longtable}

\begin{longtable}[c]{|c||c||c|}
\hline
\multicolumn{3}{|c|}{$\chi$SF} \\ \hline
\multicolumn{1}{|c||}{} &
\multicolumn{1}{|c||}{$O_{12}$} &
\multicolumn{1}{|c|}{$O_{44}$} \\ \hline
\multicolumn{3}{|c|}{Intermediate scale: $\overline{g}^{2}=2.4484$} \\\hline
$\sigma_{\mathrm{O}}(2,u)$ & 0.766\,(35)  & 0.779\,(42)    \\
slope                &  2\,(3) & -2\,(4) \\
$\chi^{2}/\mathrm{dof}$  & 1.4081         & 0.0421           \\
\hline
\multicolumn{3}{|c|}{Perturbative scale: $\overline{g}^{2}=0.9944$} \\ \hline
$\sigma_{\mathrm{O}}(2,u)$ & 0.970\,(16)    & 0.989\,(19)    \\ 
slope                   & -1\,(1) & -4\,(2) \\
$\chi^{2}/\mathrm{dof}$  & 3.3364           & 0.2731           \\
\hline
\caption{Continuum limit of the SSF, $\sigma_{\mathrm{O}_{12}}$ and
  $\sigma_{\mathrm{O}_{44}}$, of the operators $O_{12}$ and $O_{44}$.
  Results are shown for the $\chi$SF with standard Wilson fermions
  at two values of the renormalization scale and for $\boldsymbol{\theta}=(0.5,0.5,0.5)$.
  These results correspond to linear fits of the values in
  tab.~\ref{tab:LSSFO12O44Theta0.5}.
  The fits are linear in $(a/L)^{2}$.}
\label{tab:SSFcont.O12O44Theta0.5}
\end{longtable}
\newpage
\begin{longtable}[c]{|c||c|c|c||c|c|c|}
\hline
\multicolumn{1}{|c||}{} &
\multicolumn{3}{|c||}{$O_{12}$} &
\multicolumn{3}{|c|}{$O_{44}$} \\ \hline
\multicolumn{1}{|c||}{} &
\multicolumn{1}{|c|}{$\chi$SF} &
\multicolumn{1}{|c|}{SF (Clover)}  &
\multicolumn{1}{|c||}{SF (Wil)} &
\multicolumn{1}{|c|}{$\chi$SF} &
\multicolumn{1}{|c|}{SF (Clover)}  &
\multicolumn{1}{|c|}{SF (Wil)} \\\hline
\multicolumn{7}{|c|}{Intermediate scale: $\overline{g}^{2}=2.4484$} \\ \hline
$\sigma_{\mathrm{O}}(2,u)$ & 0.778\,(20)   &0.790\,(19) & 0.822\,(18) & 0.837\,(23)    &  0.812\,(23) & 0.774\,(25) \\
slope                                & 1\,(2)            &0.25\,(19)   & 0.47\,(19)   & -4\,(2) &-0.19\,(23)   & 0.17\,(26)   \\
$\chi^{2}/\mathrm{dof}$                  & 0.0772          &0.8338      & 0.0332         & 2.9018           &   0.2471           & 0.1586       \\
\hline
\multicolumn{7}{|c|}{Perturbative scale: $\overline{g}^{2}=0.9944$} \\ \hline
$\sigma_{\mathrm{O}}(2,u)$ &   0.9752\,(97) & & &   0.988\,(10) & & \\
slope                   & -1.89\,(82)     & & & -3.40\,(88)   & & \\
$\chi^{2}/\mathrm{dof}$   &  2.9783          & & &   0.0534        & & \\
\hline
\caption{Continuum limit of the SSF, $\sigma_{\mathrm{O}_{12}}$ and
  $\sigma_{\mathrm{O}_{44}}$, of the operators $O_{12}$ and $O_{44}$.
  Results are shown for the $\chi$SF with standard Wilson fermions and
  also for the SF with improved and standard Wilson fermions
  at two values of the renormalization scale and for $\boldsymbol{\theta}=(1,0,0)$.
  These results correspond to linear fits of the values in
  tab.~\ref{tab:LSSFO12O44Theta100}.
  The fits are linear in $(a/L)^{2}$ for the $\chi$SF formulation
  while they are linear in $a/L$ for the SF.}
\label{tab:SSFcont.O12O44Theta100}
\end{longtable}

\newpage
\clearpage
\section{Tables of numerical results for the renormalization factors}
\label{app:Z}
\scriptsize

\begin{minipage}[t]{0.5\linewidth}
\begin{longtable}[c]{|r|l|l|}\hline
\multicolumn{1}{|c|}{$L/a$} &
\multicolumn{1}{|c|}{$\beta$} &
\multicolumn{1}{|c|}{$Z_{\mathrm{P}}$ } \\
\hline
  8 & 6.0219 & 0.5385\,(12) \\
10 & 6.1628 & 0.5264\,(12) \\
12 & 6.2885 & 0.5272\,(16) \\
16 & 6.4956 & 0.5187\,(22) \\
\hline
\caption{$Z_{\mathrm{P}}(g_{0},L/a)$ at $L=1.436\, r_{0}$ and for $\boldsymbol{\theta}=(0.5,0.5,0.5)$.}
\label{tab:ZPmatching.Theta0.5}
\end{longtable}
\end{minipage}
\hspace{1.0cm}
\begin{minipage}[t]{0.5\linewidth}
\begin{longtable}[c]{|r|r @{.} l|r @{.} l|}\hline
\multicolumn{1}{|c|}{$\mathrm{i}$} &
\multicolumn{2}{|c|}{$z_{i}^{\textrm{P}}$} &
\multicolumn{2}{|c|}{$z_{i}^{\textrm{M}}$} \\
\hline
0 &   0&5394\,(14) &   2&1444\,(55)\\
1 & -0&077\,(15)   &   0&321\,(60)\\
2 &   0&078\,(30)   & -0&32\,(12)\\
\hline
\caption{Coefficients of the beta dependence of
  $Z_{\mathrm{P}}(g_{0},L/a)$
  at the matching scale $L=1.436\,r_{0}$
  (cf. Eq.~\eqref{eq:ZPmatchingbeta})
  and $Z_{\mathrm{M}}(g_{0})$ (cf. Eq.~\eqref{eq:ZMbeta}).
  Results are shown for the $\chi$SF with standard Wilson fermions
  at $\boldsymbol{\theta}=(0.5,0.5,0.5)$.}
\label{tab:ZPbetaNP}
\end{longtable}
\end{minipage}

\begin{minipage}[t]{0.5\linewidth}
\begin{longtable}[c]{|r|l|l|l|}\hline
\multicolumn{1}{|c|}{$L/a$} &
\multicolumn{1}{|c|}{$\beta$} &
\multicolumn{1}{|c|}{$Z_{\mathrm{P}}^{-1}$} &
\multicolumn{1}{|c|}{$Z_{\mathrm{M}}$ } \\
\hline
8   & 6.0219 & 1.8569\,(41) & 2.1484\,(47) \\
10 & 6.1628 & 1.8995\,(42) & 2.1977\,(49) \\
12 & 6.2885 & 1.8968\,(59) & 2.1946\,(68) \\
16 & 6.4956 & 1.9279\,(80) & 2.2306\,(93) \\
\hline
\caption{$Z_{\mathrm{P}}^{-1}(g_{0},L/a)$ at $L=1.436\, r_{0}$
  and $Z_{\mathrm{M}}(g_{0})$, for $\boldsymbol{\theta}=(0.5,0.5,0.5)$.
  Results are shown for the $\chi$SF formulation and for all the
  values of $\beta$ where simulations have been performed.}
\label{tab:ZMmatching}
\end{longtable}
\end{minipage}
\hspace{1.0cm}
\begin{minipage}[t]{0.5\linewidth}
\begin{longtable}[c]{|l|l|l|}\hline
\multicolumn{1}{|c|}{$\beta$} &
\multicolumn{1}{|c|}{$Z_{\mathrm{P}}$ } &
\multicolumn{1}{|c|}{$Z_{\mathrm{M}}$ } \\
\hline
6.00 & 0.5394\,(14)   & 2.1444\,(55)\\
6.10 & 0.53240\,(82) & 2.1733\,(33)\\
6.20 & 0.5270\,(10)   & 2.1957\,(42)\\
6.45 & 0.5203\,(17)   & 2.2236\,(70)\\
\hline
\caption{$Z_{\mathrm{P}}(g_{0},L/a)$ at $L=1.436\, r_{0}$
  and $Z_{\mathrm{M}}(g_{0})$, for $\boldsymbol{\theta}=(0.5,0.5,0.5)$.
  Results are presented for the $\chi$SF formulation at several
  values of $\beta$, as determined from the curves in
  eq.~\eqref{eq:ZPmatchingbeta}
  and eq.~\eqref{eq:ZMbeta}.}
\label{tab:ZMbetainterest}
\end{longtable}
\end{minipage}

\begin{longtable}[c]{|r||r @{.} l|r @{.} l|r @{.} l||r @{.} l|r @{.} l|r @{.} l|}
\hline
\multicolumn{1}{|c||}{} &
\multicolumn{6}{|c||}{$z_{i}(O_{12})$} &
\multicolumn{6}{|c|}{$z_{i}(O_{44})$} \\\hline
\multicolumn{1}{|c||}{$\mathrm{i}$} &
\multicolumn{2}{|c|}{$\chi$SF} &
\multicolumn{2}{|c|}{SF (Clover)} &
\multicolumn{2}{|c||}{SF (Wilson)} &
\multicolumn{2}{|c|}{$\chi$SF} &
\multicolumn{2}{|c|}{SF (Clover)} &
\multicolumn{2}{|c|}{SF (Wilson)} \\ \hline
\multicolumn{13}{|c|}{$\boldsymbol{\theta}=(0.5,0.5,0.5)$} \\\hline
0 &   0&402\,(14) & \multicolumn{2}{|c|}{} & \multicolumn{2}{|c||}{} &0&323\,(12) & \multicolumn{2}{|c|}{} & \multicolumn{2}{|c|}{} \\
1 & -0&25\,(15)   & \multicolumn{2}{|c|}{} & \multicolumn{2}{|c||}{} & -0&12\,(12) & \multicolumn{2}{|c|}{} & \multicolumn{2}{|c|}{} \\
2 &   0&30\,(29)   & \multicolumn{2}{|c|}{} & \multicolumn{2}{|c||}{} & -0&02\,(25) & \multicolumn{2}{|c|}{} & \multicolumn{2}{|c|}{} \\
\hline
\multicolumn{13}{|c|}{$\boldsymbol{\theta}=(1,0,0)$} \\\hline
0 &   0&3761\,(69) &   0&3410\,(31) &   0&3659\,(35) &   0&3426\,(73) &   0&3450\,(37) &   0&3197\,(44)\\
1 & -0&151\,(72) & -0&077\,(31) & -0&102\,(35) & -0&059\,(77) & -0&180\,(37) & -0&117\,(44)\\
2 &   0&18\,(15) &   0&061\,(62) &   0&047\,(70) & -0&07\,(16) &   0&196\,(72) &   0&105\,(89)\\
\hline
\caption{Coefficients of the beta-dependence of $Z_{\mathrm{O}_{12}}$
  and $Z_{\mathrm{O}_{44}}$ at the matching scale $L=1.436\,r_{0}$.
  Results are shown for the $\chi$SF with standard Wilson fermions
  at $\boldsymbol{\theta}=(1,0,0)$ and $\boldsymbol{\theta}=(0.5,0.5,0.5)$
  and also for the SF~\protect{\cite{Guagnelli:2004ga}} with improved and
  standard Wilson fermions at $\boldsymbol{\theta}=(1,0,0)$.}
\label{tab:ZObetaNP}
\end{longtable}
\newpage
\begin{longtable}[c]{|r|l||l|l|l||l|l|l|}
\hline
\multicolumn{2}{|c||}{} &
\multicolumn{3}{|c||}{$Z_{\textrm{O}_{12}}^{\mathrm{RGI}}(g_{0})$} &
\multicolumn{3}{|c|}{$Z_{\textrm{O}_{44}}^{\mathrm{RGI}}(g_{0})$}\\
\hline
\multicolumn{1}{|c|}{$L/a$} &
\multicolumn{1}{|c||}{$\beta$} &
\multicolumn{1}{|c|}{$\chi$SF} &
\multicolumn{1}{|c|}{SF (Clo)} &
\multicolumn{1}{|c||}{SF (Wil)} &
\multicolumn{1}{|c|}{$\chi$SF} &
\multicolumn{1}{|c|}{SF (Clo)} &
\multicolumn{1}{|c|}{SF (Wil)} \\
\hline
  8&6.0219&1.548\,(24)&1.414\,(13)&1.518\,(14)&1.546\,(28)&1.562\,(17)&1.453\,(20)\\
10&6.1628&1.450\,(24)&1.347\,(13)&1.433\,(14)&1.495\,(28)&1.450\,(16)&1.355\,(19)\\
12&6.2885&1.466\,(32)&1.358\,(12)&1.431\,(14)&1.456\,(39)&1.417\,(16)&1.361\,(19)\\
16&6.4956&1.409\,(45)&1.309\,(15)&1.348\,(18)&1.344\,(54)&1.371\,(19)&1.295\,(26)\\
\hline
\caption{RGI renormalization factors,
  $Z_{\textrm{O}_{12}}^{\mathrm{RGI}}(g_{0})$ and
  $Z_{\textrm{O}_{44}}^{\mathrm{RGI}}(g_{0})$ for $\boldsymbol{\theta}=(1,0,0)$.
  Results are shown for the $\chi$SF with standard Wilson fermions
  and the SF with improved and standard Wilson fermions and
  for several values of the lattice spacing.
  We have determined in this work the RGI
  Z-factors for the SF from the Z-factors given in~\protect{\cite{Guagnelli:2004ga}}.}
\label{tab:ZORGINPTheta100}
\end{longtable}

\begin{longtable}[c]{|r||r @{.} l|r @{.} l|r @{.} l||r @{.} l|r @{.} l|r @{.} l|}
\hline
\multicolumn{1}{|c||}{} &
\multicolumn{6}{|c||}{$z_{i}^{\mathrm{RGI}}(O_{12})$} &
\multicolumn{6}{|c|}{$z_{i}^{\mathrm{RGI}} (O_{44})$} \\
\hline
\multicolumn{1}{|c||}{$\mathrm{i}$} &
\multicolumn{2}{|c|}{$\chi$SF} &
\multicolumn{2}{|c|}{SF (Clover)} &
\multicolumn{2}{|c||}{SF (Wilson)} &
\multicolumn{2}{|c|}{$\chi$SF} &
\multicolumn{2}{|c|}{SF (Clover)} &
\multicolumn{2}{|c|}{SF (Wilson)} \\
\hline
0 &  1&554\,(29) &  1&409\,(13) &  1&512\,(14)&  1&550\,(33)&  1&561\,(17)&  1&447\,(20)\\
1 &-0&62\,(30)   &-0&32\,(13)   &-0&42\,(14)  &-0&27\,(35)  &-0&81\,(17)  &-0&53\,(20)\\
2 &  0&74\,(62)   &  0&25\,(26)   &  0&19\,(29)  &-0&32\,(72)  &  0&89\,(33)  &  0&48 \,(40)\\
\hline
\caption{Coefficients of the beta-dependence of
  $Z_{\mathrm{O}_{12}}^{\mathrm{RGI}}(g_{0})$
  and $Z_{\mathrm{O}_{44}}^{\mathrm{RGI}}(g_{0})$.
  Results are shown for the $\chi$SF with standard Wilson fermions
  and the SF with improved and standard Wilson
  fermions at $\boldsymbol{\theta}=(1,0,0)$.} 
\label{tab:ZORGIbetaNP}
\end{longtable}

\begin{longtable}[c]{|l||l|l||l|l|}
\hline
\multicolumn{1}{|c||}{$\beta$} &
\multicolumn{1}{|c|}{$Z_{\mathrm{O}_{12}}$} &
\multicolumn{1}{|c||}{$Z_{\mathrm{O}_{12}}^{\mathrm{RGI}}$} &
\multicolumn{1}{|c|}{$Z_{\mathrm{O}_{44}}$} &
\multicolumn{1}{|c|}{$Z_{\mathrm{O}_{44}}^{\mathrm{RGI}}$} \\
\hline
\multicolumn{1}{|c||}{} &
\multicolumn{4}{|c|}{$\chi$SF} \\
\hline
6.00 & 0.3761\,(69) & 1.554\,(29) & 0.3426\,(73) & 1.550\,(33)\\ 
6.10 & 0.3627\,(40) & 1.499\,(17) & 0.3361\,(43) & 1.521\,(19)\\
6.20 & 0.3529\,(49) & 1.458\,(20) & 0.3283\,(53) & 1.486\,(24)\\
6.45 & 0.3436\,(82) & 1.420\,(34) & 0.3031\,(90) & 1.371\,(41)\\
\hline
\multicolumn{1}{|c||}{} &
\multicolumn{4}{|c|}{SF (Clover)} \\
\hline
6.00&0.3410\,(31)&1.409\,(13)  &0.3450\,(37)&1.561\,(17)\\
6.10&0.3338\,(20)&1.3793\,(83)&0.3290\,(23)&1.489\,(10)\\
6.20&0.3279\,(23)&1.3550\,(95)&0.3168\,(27)&1.433\,(12)\\
6.45&0.3186\,(30)&1.317\,(12)  &0.3035\,(35)&1.373\,(16)\\
\hline
\multicolumn{1}{|c||}{} &
\multicolumn{4}{|c|}{SF (Wilson)} \\
\hline
6.00&0.3659\,(35)&1.512\,(14)  &0.3197\,(44)&1.447\,(20)\\
6.10&0.3562\,(22)&1.4719\,(91)&0.3091\,(28)&1.399\,(13)\\
6.20&0.3474\,(25)&1.436\,(10)  &0.3005\,(32)&1.360\,(14)\\
6.45&0.3296\,(36)&1.362\,(15)  &0.2884\,(46)&1.305\,(21)\\
\hline
\caption{$Z_{\mathrm{O}}(g_{0},L/a)$ at scale $L=1.436\, r_{0}$
  and
  $Z_{\mathrm{O}}^{\mathrm{RGI}}(g_{0})$
  for both $O_{12}$ and $O_{44}$ and for $\boldsymbol{\theta}=(1,0,0)$.
  Results are presented for the $\chi$SF with standard Wilson fermions
  and the SF with improved and standard Wilson fermions.}
\label{tab:ZObetainterest}
\end{longtable}

\newpage
\end{appendix}
\newpage
\bibliographystyle{h-elsevier}    
\bibliography{paper2}      

\begin{thebibliography}{10}

\bibitem{Lopez:2012as}
J.G. Lopez et~al.,
\newblock (2012), 1208.4591.

\bibitem{Luscher:1992an}
M. Luscher et~al.,
\newblock Nucl. Phys. B384 (1992) 168, hep-lat/9207009.

\bibitem{Sint:1993un}
S. Sint,
\newblock Nucl. Phys. B421 (1994) 135, hep-lat/9312079.

\bibitem{Luscher:2006df}
M. Luscher,
\newblock JHEP 05 (2006) 042, hep-lat/0603029.

\bibitem{Luscher:1993gh}
M. Luscher et~al.,
\newblock Nucl. Phys. B413 (1994) 481, hep-lat/9309005.

\bibitem{Capitani:1998mq}
ALPHA, S. Capitani et~al.,
\newblock Nucl. Phys. B544 (1999) 669, hep-lat/9810063.

\bibitem{Guagnelli:2003hw}
Zeuthen-Rome / ZeRo, M. Guagnelli et~al.,
\newblock Nucl. Phys. B664 (2003) 276, hep-lat/0303012.

\bibitem{Pena:2004gb}
C. Pena, S. Sint and A. Vladikas,
\newblock JHEP 09 (2004) 069, hep-lat/0405028.

\bibitem{DellaMorte:2004bc}
ALPHA, M. Della~Morte et~al.,
\newblock Nucl. Phys. B713 (2005) 378, hep-lat/0411025.

\bibitem{DellaMorte:2005kg}
ALPHA, M. Della~Morte et~al.,
\newblock Nucl. Phys. B729 (2005) 117, hep-lat/0507035.

\bibitem{Aoki:2009tf}
PACS-CS Collaboration, S. Aoki et~al.,
\newblock JHEP 0910 (2009) 053, 0906.3906.

\bibitem{Aoki:2010wm}
PACS-CS collaboration, S. Aoki et~al.,
\newblock JHEP 1008 (2010) 101, 1006.1164.

\bibitem{Tekin:2010mm}
ALPHA, F. Tekin, R. Sommer and U. Wolff,
\newblock Nucl. Phys. B840 (2010) 114, 1006.0672.

\bibitem{Frezzotti:2003ni}
R. Frezzotti and G.C. Rossi,
\newblock JHEP 08 (2004) 007, hep-lat/0306014.

\bibitem{Sint:2010eh}
S. Sint,
\newblock Nucl. Phys. B847 (2011) 491, 1008.4857.

\bibitem{Lopez:2008ns}
J.G. Lopez, K. Jansen and A. Shindler,
\newblock PoS LATTICE2008 (2008) 242, 0810.0620.

\bibitem{Lopez:2009yc}
J.G. Lopez et~al.,
\newblock PoS LAT2009 (2009) 199, 0910.3760.

\bibitem{Sint:2010xy}
S. Sint and B. Leder,
\newblock PoS LATTICE2010 (2010) 265, 1012.2500.

\bibitem{Boucaud:2007uk}
ETM, P. Boucaud et~al.,
\newblock (2007), hep-lat/0701012.

\bibitem{Guagnelli:2004za}
ALPHA, M. Guagnelli et~al.,
\newblock JHEP 05 (2004) 001, hep-lat/0402022.

\bibitem{Bucarelli:1998mu}
A. Bucarelli et~al.,
\newblock Nucl. Phys. B552 (1999) 379, hep-lat/9808005.

\bibitem{Frezzotti:2000nk}
ALPHA, R. Frezzotti et~al.,
\newblock JHEP 08 (2001) 058, hep-lat/0101001.

\bibitem{Guagnelli:2004ga}
Zeuthen-Rome (ZeRo), M. Guagnelli et~al.,
\newblock Eur. Phys. J. C40 (2005) 69, hep-lat/0405027.

\bibitem{Guagnelli:1999wp}
M. Guagnelli, K. Jansen and R. Petronzio,
\newblock Phys.Lett. B459 (1999) 594, hep-lat/9903012.

\bibitem{Capitani:2005jp}
S. Capitani et~al.,
\newblock Phys. Lett. B639 (2006) 520, hep-lat/0511013.

\bibitem{Sommer:1993ce}
R. Sommer,
\newblock Nucl. Phys. B411 (1994) 839, hep-lat/9310022.

\bibitem{Garden:1999fg}
ALPHA, J. Garden et~al.,
\newblock Nucl. Phys. B571 (2000) 237, hep-lat/9906013.

\bibitem{Jansen:2005kk}
XLF, K. Jansen et~al.,
\newblock JHEP 09 (2005) 071, hep-lat/0507010.

\bibitem{Guagnelli:1998ud}
ALPHA, M. Guagnelli, R. Sommer and H. Wittig,
\newblock Nucl. Phys. B535 (1998) 389, hep-lat/9806005.

\bibitem{Shindler:2007vp}
A. Shindler,
\newblock Phys. Rept. 461 (2008) 37, 0707.4093.

\bibitem{Gasser:1982ap}
J. Gasser and H. Leutwyler,
\newblock Phys. Rept. 87 (1982) 77.

\bibitem{Leutwyler:1994fi}
H. Leutwyler,
\newblock (1994), hep-ph/9406283.

\bibitem{Leutwyler:1996qg}
H. Leutwyler,
\newblock Phys. Lett. B378 (1996) 313, hep-ph/9602366.

\bibitem{Martinelli:1994ty}
G. Martinelli et~al.,
\newblock Nucl. Phys. B445 (1995) 81, hep-lat/9411010.

\bibitem{Constantinou:2010gr}
ETM, M. Constantinou et~al.,
\newblock JHEP 08 (2010) 068, 1004.1115.

\bibitem{Gimenez:2004me}
V. Gimenez et~al.,
\newblock Phys.Lett. B598 (2004) 227, hep-lat/0406019.

\bibitem{Cichy:2012is}
K. Cichy, K. Jansen and P. Korcyl,
\newblock (2012), 1207.0628.

\bibitem{Jansen:1995ck}
K. Jansen et~al.,
\newblock Phys.Lett. B372 (1996) 275, hep-lat/9512009.

\end{thebibliography}
\end{document}